\documentclass[lettersize,journal]{IEEEtran}
\usepackage{amsmath,amsfonts}
\usepackage{algorithm}
\usepackage{array}
\usepackage[caption=false,font=normalsize,labelfont=sf,textfont=sf]{subf    ig}
\usepackage{textcomp}
\usepackage{stfloats}
\usepackage{url}
\usepackage{verbatim}
\usepackage{graphicx}
\usepackage{cite}
\usepackage{wrapfig}
\usepackage{booktabs}
\usepackage{multirow}
\usepackage{makecell}
\usepackage{caption}
\usepackage{xcolor}
\usepackage{amsmath}
\usepackage{float}
\usepackage{amssymb}
\usepackage{algpseudocode}
\usepackage{algorithm}
\usepackage{amsmath}
\usepackage{flushend}
\usepackage{wrapfig}
\usepackage[breaklinks=true]{hyperref}
\usepackage{url}

\makeatletter
\def\@IEEEBIOskipN{1\baselineskip}
\makeatother

\begin{document}
\raggedbottom
\title{Doppler-Domain Respiratory Amplification for
Semi-Static Human Occupancy Detection Using
Low-Resolution SIMO FMCW Radar}

\author{Huy Trinh, \IEEEmembership{Member, IEEE}, 
Elliot Creager, and George Shaker, \IEEEmembership{Senior Member, IEEE}
\thanks{This paper was produced by the IEEE Publication Technology Group. They are in Piscataway, NJ.}
\thanks{Manuscript received April 19, 2021; revised August 16, 2021.}}

\markboth{Journal of \LaTeX\ Class Files,~Vol.~14, No.~8, August~2021}%
{Shell \MakeLowercase{\textit{et al.}}: A Sample Article Using IEEEtran.cls for IEEE Journals}

\IEEEpubid{0000--0000/00\$00.00~\copyright~2021 IEEE}

\maketitle

\begin{abstract}
Radar-based sensing is a promising privacy-preserving alternative to cameras and wearables in settings such as long-term care. Yet detecting quasi-static presence (lying, sitting, or standing with only subtle micro-motions) is difficult for low-resolution SIMO FMCW radar because near-zero Doppler energy is often buried under static clutter. We present Respiratory-Amplification Semi-Static Occupancy (RASSO), an invertible Doppler-domain non-linear remapping that densifies the slow-time FFT (Doppler) grid around 0 m/s before adaptive Capon beamforming. The resulting range-azimuth (RA) maps exhibit higher effective SNR, sharper target peaks, and lower background variance, making thresholding and learning more reliable. On a real nursing-home dataset collected with a short-range 1Tx-3Rx radar, RASSO-RA improves classical detection performance, achieving AUC = 0.981 and recall = 0.920/0.947 at FAR = 1\%/5\%, outperforming conventional Capon processing and a recent baseline. RASSO-RA also benefits data-driven models: a frame-based CNN reaches 95-99\% accuracy and a sequence-based CNN-LSTM reaches 99.4-99.6\% accuracy across subjects. A paired session-level bootstrap test confirms statistically significant macro-F1 gains of 2.6-3.6 points (95\% confidence intervals above zero) over the non-warped pipeline. These results show that simple Doppler-domain warping before spatial processing can materially improve semi-static occupancy detection with low-resolution radar in real clinical environments.

\end{abstract}


\section{Introduction}
\label{sec:introduction}
\IEEEPARstart{R}{adar} technology has a long history of application in various sectors, including aerospace, defence, and automotive systems, with the recent emerging field of radar-based human activity recognition (HAR) \cite{9704291}, \cite{10012054}. Unlike cameras or wearable devices, radar offers a contactless and privacy-preserving solution for monitoring well-being. However, in practice, the reliability of radar is challenged by the need to detect senior and elderly people, who most of the time are in quasi-static states or perform subtle movements. These scenarios produce only faint micro-Doppler ($\mu$-Doppler) signatures from respiration, which can hide weak reflections and raise the background floor, therefore hindering precise activity detection. Moreover, many present studies utilize low-resolution radar for occupancy studies and other applications \cite{10784889}, \cite{10880536}. The main contributions of our research can be summarized in two points:
\begin{itemize}
    \item We propose RASSO, an invertible Doppler-domain non-linear warping and resampling of the slow-time FFT axis that densifies the Doppler sampling grid around near-zero micro-motions before Capon beamforming, yielding range–Azimuth maps with higher signal concentration and output signal-to-noise ratio.
    \item We evaluate the impact of RASSO on real datasets collected at the nursing home setting using both knowledge-driven and data-driven approaches. RASSO demonstrate high accuracy generalization across subjects and statistically significant improvements under uncertainty quantification.
\end{itemize}

The rest of the paper is organized as follows. Section~\ref{sec:related_works} summarizes existing research works in quasi-static detection using FMCW and other radar types. Section~\ref{sec:methodology} gives an overview of the technical background, preprocessing pipeline and proposes the RASSO algorithm in more detail. Section~\ref{sec:result} presents our experiment setup against the vendor's standard beamforming method and other peers' research work, reports results for (i) knowledge-driven CFAR detection for cross-subject, (ii) data-driven occupancy classification and bootstrapping for uncertainty quantification. Finally, Section~\ref{sec:conclusion} discusses the implications and concludes the study.
\section{RELATED WORK}
\label{sec:related_works}
Reliable indoor quasi-static occupant detection has long been known to be challenging due to its very low Doppler shift; even recent FMCW systems acknowledge that detecting static people is more difficult than detecting moving targets \cite{10118759}, \cite{book}. Early efforts to detect stationary occupants focused on resolving respiratory motion using ultra-wideband (UWB) or continuous wave (CW) radars. For example, Song et al. \cite{bios15050273} developed an algorithm to detect a stationary person based on the spectrum of a demodulated respiration signal, specifically targeted breathing frequencies. This approach requires the subject's respiration to be sensitive to distance-null points. Additionally, FMCW radars provide range and angle information, enabling indoor localization of people. A study by Luo et al. \cite{9951449} proposed a standard-deviation weighting method to locate a person by emphasizing range bins with high amplitude variance across chirps, but their system still relies on heuristic weighting. Later, Kaiser et al. \cite{10289261} improved sem-static human localization by combining range-angle estimates with clutter suppression and thresholding. In parallel, Li et al. \cite{10118759} employed Multiple Signal Classification (MUSIC) processing with Multiple-Input Multiple-Output (MIMO) arrays to localize standing subjects, but their approach requires multiple antennas and is computationally intensive. These methods generally treat stationary detection as a by-product of localization and do not explicitly address the faint micro-Doppler components of semi-static subjects. 
 
Sabri et al. \cite{kahya2023mcroodmulticlassradaroutofdistribution} introduce the MCROOD framework operates on Range Doppler (RD) images to detect reconstruction-based out-of-distribution (OOD). In other words, it simultaneously predicts whether a human is present and flags unfamiliar disturbances. Their method relies on E-RESPD, which temporally accumulated a heatmap produced by sliding-window summation/averaging over magnitude RDIs. While promising, E-RESPD's consecutive-frame averaging methods eliminate per-antenna phase, which cannot be used directly for angle-of-arrival (AoA) beamforming. In addition, like many previous works, their work mainly reports results from controlled laboratory environments and has not yet been shown to deploy in dynamic scenarios. To ensure compliance, we also strive to reproduce their work and compare it fairly with our methods, as shown later in Section \ref{sec:result}. To resolve these issues and overcome these disadvantages, we introduce RASSO, an invertible Doppler-domain non-linear warping and resampling of the Doppler axis that densifies the sampling grid around near-zero micro-Doppler before spatial processing, thereby preserving per-antenna phase while making faint $\mu$-Doppler signatures more prominent in the subsequent RA maps.
Integrated with Minimum Variance Distortionless Response (MVDR)/Capon beamforming and 2-D CFAR, our signal-processing pipeline boosts the SNR of quasi-static targets and enables a compact CNN/LSTM classifier to reach $\approx$ 92-99\% accuracy on both held-out and generalization tests in realistic living-room scenarios. Unlike previous work, our results emphasize robustness in real-world settings with clutter and semi-static posture changes, supporting dependable deployment beyond the lab. 
\section{METHODOLOGY}
\label{sec:methodology}
\subsection{Experiment Setup and Data Collection}
\begin{figure}[H]
\centerline{\includegraphics[width=\linewidth]{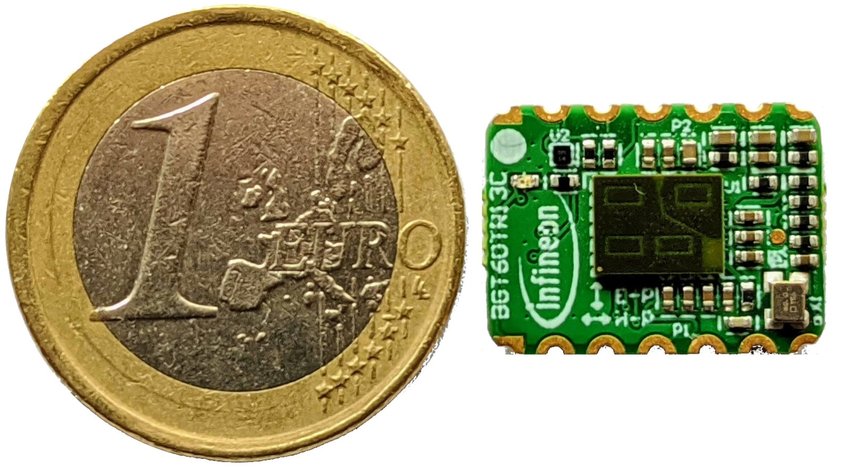}}
\caption{Infineon XENSIV\textsuperscript{\texttrademark} BGT60TR13C Radar. \cite{infineon2024bgt60tr13c} }
\label{fig:bgt60_image}
\end{figure}

\begin{figure}[H]
\centerline{\includegraphics[width=\linewidth]{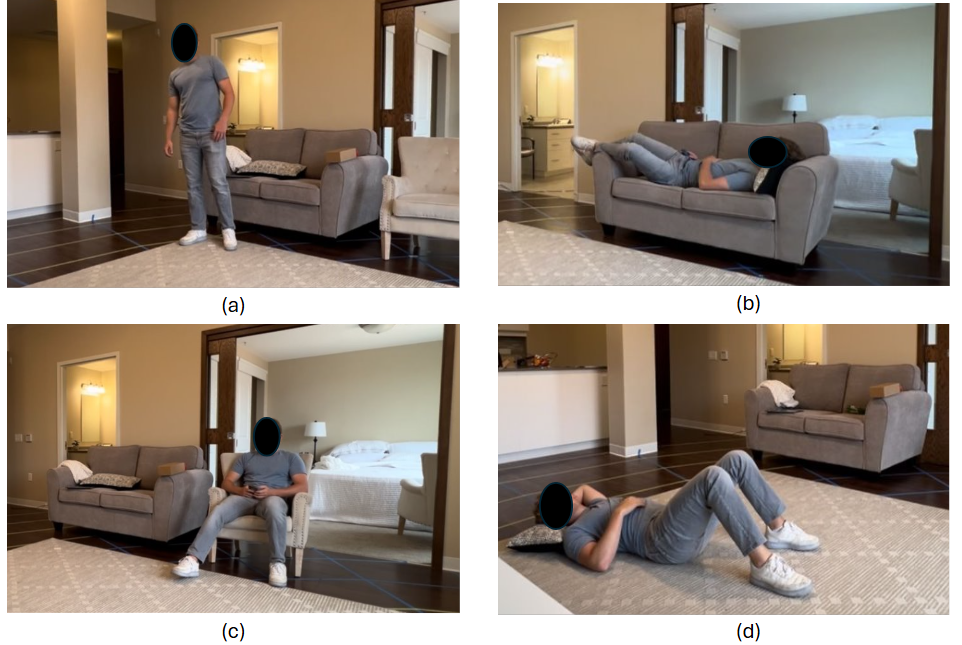}}
\caption{Few measurement activities: Standing (a), Lay on Sofa (b), Sitting (c), Lay on Floor (d).}
\label{fig:experiment_setup_josh}
\end{figure}

For this study, we collected a dedicated dataset of semi-static activities in a long-term care facility using an off-the-shelf 60\, GHz mmWave FMCW radar (Infineon XENSIV\textsuperscript{\texttrademark} BGT60TR13C) with one transmitter and three receivers \cite{infineon2024bgt60tr13c}. The hardware and waveform parameters used in all recordings are summarized in Table~\ref{tab:radar-hw}. Two identical radars were wall-mounted at approximately 2.5 m above the ground on different walls, with partially overlapping fields of view and a slight downward tilt to cover both the sofa and the floor areas. The room contains common furnishings (sofa, coffee table, TV, windows, and surrounding walls), inducing realistic multipath and occlusions. The two radars were wall-mounted at approximately 2.5 m above the ground on different walls, with partially overlapping fields of view and a slight downward tilt to cover both the sofa and the floor areas. This placement provided angular diversity while keeping sensor positions out of the way of daily activities. Unlike many “controlled” lab datasets, ours targets real-life long-term care facility (LTC) conditions: participants are not strictly constrained, and recordings include idle posture (lying on the floor/sofa, sitting on the floor/sofa, standing) as well as natural micro-movements (posture micro-adjustments while sitting/standing), and brief low-motion transitions (e.g., getting up from a sofa, picking an object from the floor/table) as shown in Fig.~\ref{fig:experiment_setup_josh}. For each activity, we recorded 2 minutes at 5 distinct positions and orientations relative to the sensors to vary the aspect angle, range, and background layout. Radar frames were acquired continuously at 10\, Hz, resulting in a total of approximately 384,000 frames. 
\begin{figure*}[!t] 
  \centering
  \includegraphics[
    width=\textwidth,
    keepaspectratio,          
    trim={0 0 0 0pt},        
    clip
  ]{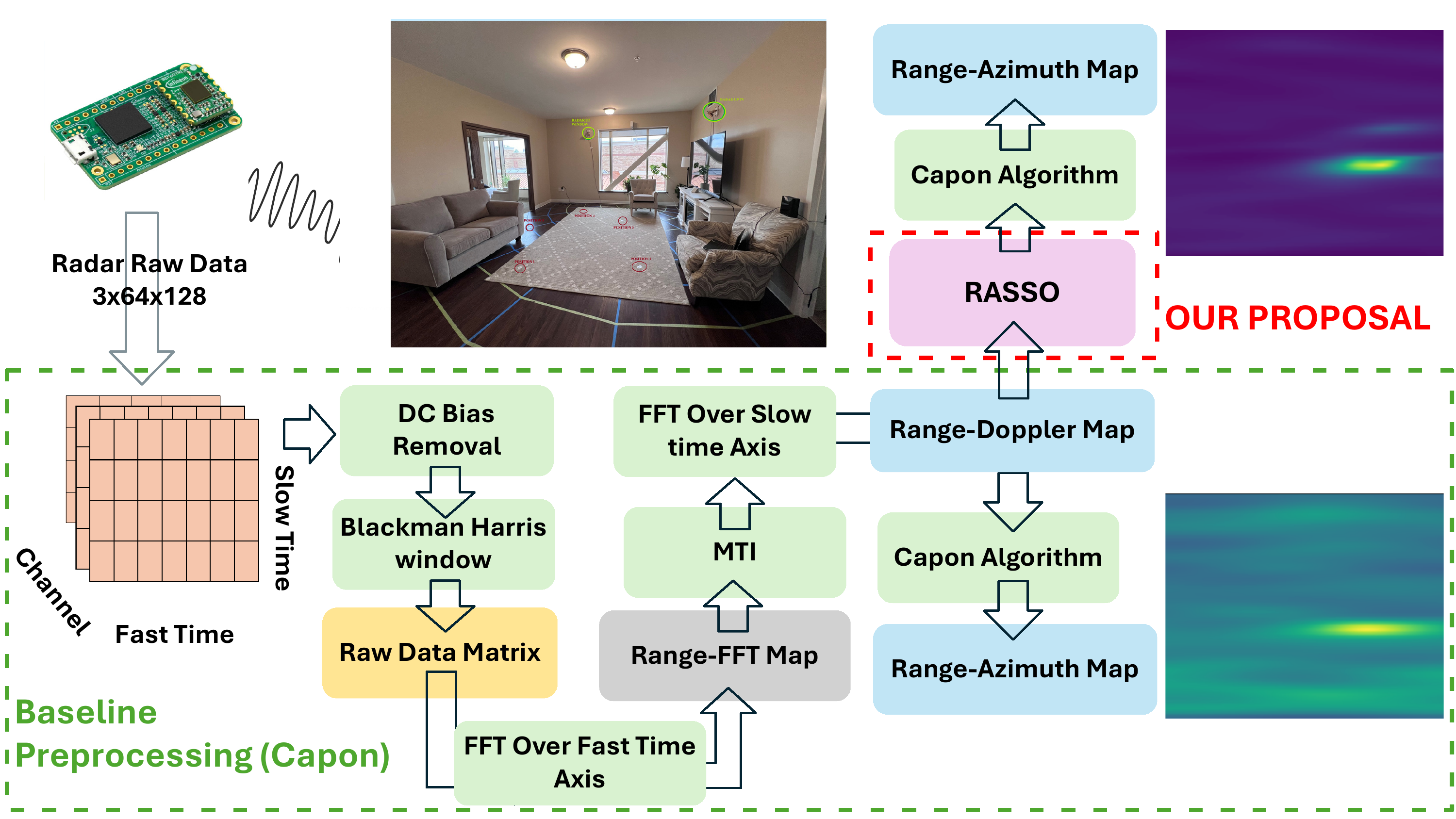}
  \caption{Our baseline signal processing pipeline (green-dash) \cite{10784889}, \cite{10880536}, \cite{Abedi2020OnTU},  and our proposed RASSO (red-dash box)}
  \label{fig:flow_chart}
\end{figure*}

\begin{table}[t]
\centering
\caption{Radar hardware and waveform configuration (BGT60TR13C).}
\label{tab:radar-hw}
\begin{tabular}{l l}
\hline
\textbf{Parameter} & \textbf{Value} \\
\hline
Band / Center frequency & 58–63.5\,GHz (center $\approx$ 60\,GHz) \\
Sweep bandwidth $B$ & $\approx$ 500\,MHz \\
Frame rate & 10\,Hz \\
Chirps per frame $N_{\text{chirp}}$ & 128 \\
Samples per chirp $N_{\text{sample}}$ & 64 \\
Range resolution $\Delta r$ & 0.30\,m \\
Max range  & 9.6\,m \\
Max unambiguous speed  & 3\,m/s \\
\hline
\end{tabular}
\end{table}

\subsection{FMCW Radar Signal Processing}
\label{subsec:preproc}

\begin{figure}[H]
\centerline{\includegraphics[trim=0.4cm 0.1cm 0.5cm 0.1cm, clip, width=\linewidth]{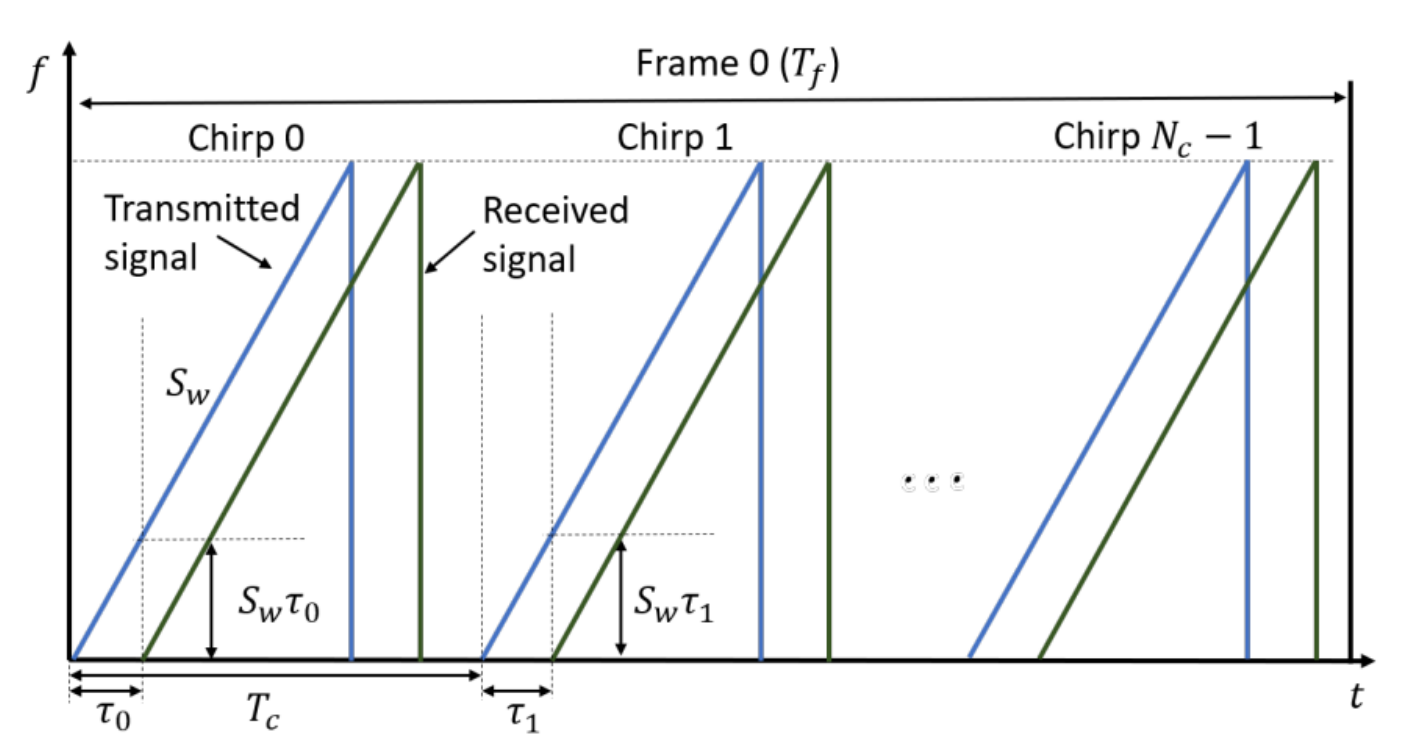}}
\caption{The transmitted FMCW chirps and corresponding return signal. \cite{phdthesis}}
\label{fig:fmcw_radar}
\end{figure}

Frequency-Modulated Continuous-Wave (FMCW) radar employs modulated continuous waves, typically operating at millimetre-wave (mmWave) frequencies, to sense the environment. It transmits a periodic wideband linear frequency-modulated, also called a chirp signal, as in Fig. \ref{fig:fmcw_radar}. This linear chirp has a duration of $T_\mathrm{c}$, bandwidth $B$, and slope \cite{TI_FMCW_Training}:

\begin{equation}
\label{eq:chirp_slope}
S_w=B/T_\mathrm{c}. 
\end{equation}

Let a point target at range $r$ induce a round–trip delay $\tau=2r/c$ (speed of light $c$).
Mixing (de-chirping) the received signal with the replica of the transmit signal (complex conjugate) produces the intermediate-frequency (IF), or \emph{beat signal}. For a single stationary target, the beat is a near-pure tone whose frequency equals chirp slope times delay \cite{TI_FMCW_Training}: 
\begin{equation}
    f_b= S_w\tau = \frac{S_w2r}{c} .
\end{equation}
The frequency of this beat signal contains information about the target's range and velocity. The basic process involves generating chirps, receiving reflections, and mixing to produce the IF signal for subsequent analysis.

\subsubsection{Range Estimation}
Range estimation relies on the time delay $\tau$ between the transmitted and received chirp signals. Due to this delay, at any given time $t$ during the chirp overlap, the instantaneous frequency difference between the transmitted and received signals is constant, which equals the IF frequency $f_{IF}$. This frequency difference is related to the delay by $f_{IF} = S \tau$ \cite{TI_FMCW_Training}. Combining these relationships, the range $d$ can be determined from the IF frequency:
\begin{equation}
    \tau = \frac{2d}{c}
    \label{eq:delay}
\end{equation}
\begin{equation}
    f_{IF} = S \tau \label{eq:if_freq_from_slope_delay}
\end{equation}
Substituting \eqref{eq:chirp_slope} and \eqref{eq:delay} into \eqref{eq:if_freq_from_slope_delay} and rearranging gives the range equation:
\begin{equation}
    d = \frac{c T_c f_{IF}}{2B}
    \label{eq1:range_formula}
\end{equation}
We first estimate the range by applying a Fast Fourier Transform (FFT) \cite{DUHAMEL1990259}, referred to as the Range-FFT, to the digitized IF signal sampled during a single chirp. Peaks in the resulting frequency spectrum indicate the presence of an object at ranges calculated using \eqref{eq1:range_formula}.
\begin{figure}[H]
\centerline{\includegraphics[width=\linewidth]{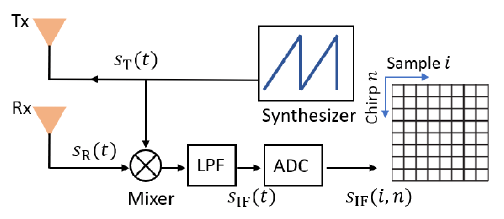}}
\caption{The receiver chain for FMCW radar. \cite{phdthesis}}
\label{fig:receiver-chain}
\end{figure}
\subsubsection{Velocity Estimation}
\begin{figure}[H]
\centerline{\includegraphics[
width=\linewidth]{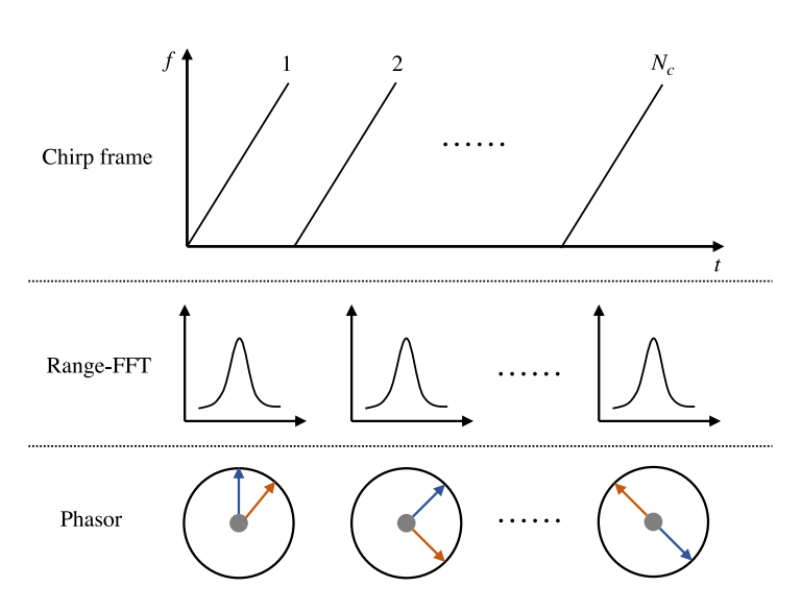}}
\caption{A chirp frame generates multiple peaks in Range-FFT, each of which has a different phase and can be utilized to estimate velocity \cite{10554983}.}
\label{fig:estimate-velocity}
\end{figure}

For a moving target with radial velocity $v$, its range changes slightly between chirps, producing a slow-time phase progression at nominal Doppler frequency $f_D = 2v/\lambda$, where $v$ is radial velocity and $\lambda$ is wavelength \cite{TI_FMCW_Training}. 
The phase $\phi$ of the IF signal is then obtained by $\phi = (2\pi \cdot 2d) / \lambda = 4\pi d / \lambda$ with round-trip path length ($2d$) and $\lambda$ is the wavelength \cite{10554983}. In general, the phase difference between two consecutive chirps separated by $T_{rep}$ is then formulated by $\Delta \phi = 4\pi \Delta d / \lambda = 4\pi v T_{rep} / \lambda$. Thus, the velocity is:
\begin{equation}
    v = \frac{\lambda \Delta \phi}{4 \pi T_{rep}}
    \label{eq:velocity_formula_simple}
\end{equation}
For a specific range, the object will induce a different phase shift $\Delta \phi_i$ across the chirps in the frame, resulting in distinct frequency components as visualized conceptually in Fig.~\ref{fig:estimate-velocity}. 

\subsubsection{Range-Doppler Map Generation}
\label{subsec:rdm_generation}
Applying Range-FFT across fast-time samples within each chirp, followed by Doppler-FFT along slow-time chirp samples for each range bin on initial ADC data cube $x[n, k]$ (samples $n$, chirps $k$) results in a two-dimensional complex-valued matrix known as the \textbf{Range-Doppler Map (RDM)}, $X_m[r, d]$. Each element $X_m[r, d]$ in the RDM represents the reflected signal strength (amplitude and phase) from a specific range bin $r$ and a specific Doppler (velocity) bin $d$. In our Single-Input and Multiple-Output  (SIMO) radar system with $M=3$ receivers, this process yields $M$ separate RDMs, $X_m[r, d]$, one for each receive antenna $m$:
\[
  X_m[r,d] \in \mathbb{C}, \quad r\in[0,N_r\!-\!1],\ d\in[0,N_d\!-\!1],\ m\in[0,N_{\mathrm{RX}}\!-\!1].
\]
Stacking these across receivers produces the \textbf{Range–Doppler Cube (RDC)},
\[
  \mathbf{Y}[r,d,m] \;=\; X_m[r,d], \qquad m=0,\ldots,N_{\mathrm{RX}}{-}1,
\]
Which supplies both inter–element phase (for digital beamforming and Capon/MVDR in Sections~\ref{subsec:dbf}–\ref{subsec:capon}) and Doppler–axis samples (for our RASSO
, Doppler-axis warping and resampling in Section~\ref{subsec:rasso}).

\paragraph*{Clutter suppression}
A critical challenge for indoor sensing is static clutter (walls, furniture, etc), which creates a massive signal at zero Doppler velocity (the DC component) in the RDM. To suppress this, we employed a \textbf{Moving Target Indicator filter} (MTI) \cite{8703820}. 

For each bin $(r,d)$ in the RDM at frame k, the clutter estimation $C_k(r,d)$ is updated through \cite{Infineon_DSP_Handout,infineon_an141319}:
\begin{equation}
    C_k(r,d) = \alpha \cdot C_{k-1}(r,d) + (1-\alpha) \cdot RDM_k(r,d)
\end{equation}
\begin{equation}
    Y_k(r,d) = RDM_k(r,d) - C_k(r, d)
\end{equation}
where $\alpha$ is the forgetting factor.

\subsection{Digital Beamforming (DBF) from Infineon}
\label{subsec:dbf}
Following the range–Doppler (RD) preprocessing and moving-target indication (MTI) described in Section~\ref {subsec:preproc}, we estimate target bearing by \emph{standard} (delay-and-sum) digital beamforming (DBF) using the vendor’s baseline implementation for the BGT60TR13C device. Particularly for our specific quasi-static application, we adapt the Infineon SDK code sample \texttt{range-angle-map.py} to our sensor configuration specified in Table \ref{tab:radar-hw}, modify MTI/static-suppression settings to be compatible with our processors for a fair comparison.\footnote{See Infineon’s DBF application notes for BGT60TR13C, we follow their geometry and weight definition. \cite{infineonAN155322,infineon_an141319}} 

\begin{figure}[H]
\centerline{\includegraphics[width=0.8\linewidth]{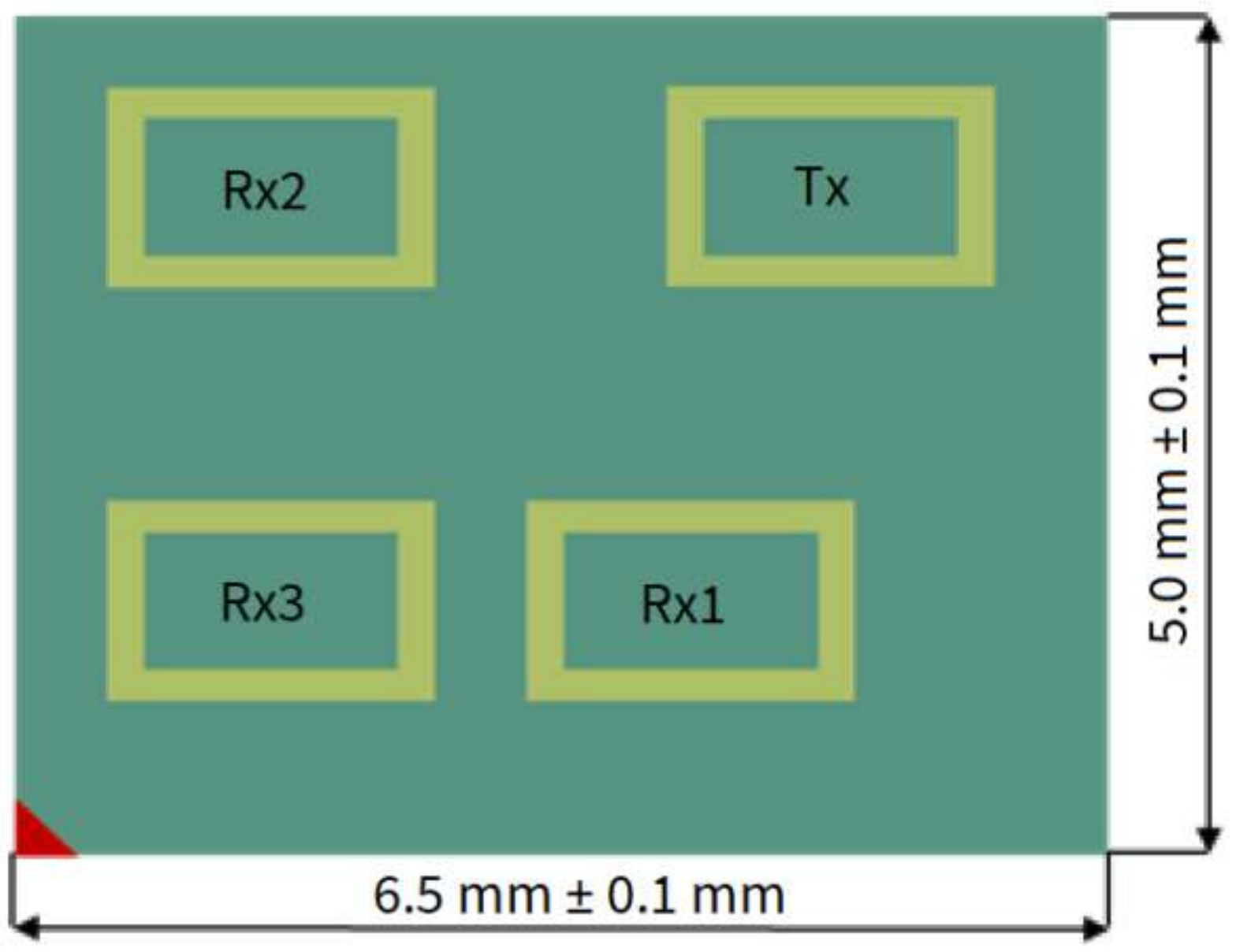}}
\caption{BGT60TR13C radar package outline \cite{infineon_an141319}.}
\label{fig:L-shape}
\end{figure}

As laid out in the Fig.~\ref{fig:L-shape}, three receivers (Rx1, Rx2, Rx3) are arranged in an L-shape to enable 2-D AoA sensing. We adopt Rx3 as the reference element and origin. With inter-element spacings $d_x=d_y=\lambda/2$, the RX element position vectors (in meters) are (\cite{infineon_an141319}, Eq.~(2)):
\begin{align}
\overrightarrow{\mathrm{Rx_1}} &= \bigl[\tfrac{\lambda}{2},\,0,\,0\bigr]^\top = [d_x,0,0]^\top, \\
\overrightarrow{\mathrm{Rx_2}} &= \bigl[0,\,\tfrac{\lambda}{2},\,0\bigr]^\top = [0,d_y,0]^\top, \\
\overrightarrow{\mathrm{Rx_3}} &= [0,\,0,\,0]^\top. 
\end{align}
Cartesian coordinates $(x,y)$ of a point at polar range~$r$ and angles $(\theta,\phi)$ follow (\cite{infineon_an141319} Eq.~(3)):
\begin{equation}
x = r \cos\theta \cos\phi, 
\qquad
y = r \sin\phi.
\end{equation}

Let $\vec{k}=\tfrac{2\pi}{\lambda}\,[\cos\theta\cos\phi, \ \sin\phi]^\top$ denote the wave-number vector along the array’s azimuth ($x$) and elevation ($y$) axes. The per-direction complex weight (steering) vector that introduces the appropriate phase shifts is (\cite{infineon_an141319} Eqs.~(4)–(5)):
\begin{equation}
\mathbf{w}(\theta,\phi) 
= 
\Bigl[
e^{j \tfrac{2\pi}{\lambda} \, d_x \cos\theta\cos\phi},\;
e^{j \tfrac{2\pi}{\lambda} \, d_y \sin\phi},\;
1
\Bigr].
\label{eq:weights}
\end{equation}

\paragraph{Delay–and–sum beamformer and range–angle map.}
Let $z_m(r,k)$ denote the complex RD value for receiver $m\!\in\!\{1,2,3\}$ at range bin $r$ and slow-time (Doppler) index $k$ after MTI. For a grid of look angles $(\theta,\phi)$, DBF computes the complex beam output by phasing and summing the channels; the corresponding power for frame $k$ is (\cite{infineon_an141319} Eq.~(6))
\begin{equation}
P_{\text{DBF}}(r,\theta,\phi;k) \;=\; \sum_{m=1}^{3} z_m(r,k)\,w_m(\theta,\phi).
\label{eq:dbf-power}
\end{equation}

Following Infineon’s SDK, the 2D‐FFT DBF yields a 4‐D spectrum
$P_{\text{DBF}}(r,\theta,\phi;k)$ (range $r$, azimuth $\theta$, elevation $\phi$, frame $k$).
For visualization and downstream CA–CFAR, we obtain the RA map by
non–coherent integration across elevation,
\begin{equation}
\label{eq:dbf-ra}
\mathrm{RA}_{\text{DBF}}(r,\theta;k)
~=~ \sum_{\phi\in\Phi} \bigl| P_{\text{DBF}}(r,\theta,\phi;k) \bigr| ,
\end{equation}
which matches Infineon’s application note and the reference implementation
(\emph{range-angle-map.py}). The Infineon's application note \cite{infineonAN155322} remarks more sophisticated methods, for example, Capon/Minimum variance distortionless response (MVDR), can sharpen the mainlobe at the cost of higher computation and additional assumptions. In this study, we use the Infineon standard DBF as a standard-provided baseline and compare it to our baseline RA processor, which adopts Capon/MVDR (described in Section~\ref {subsec:capon}) that narrows the lobe and the proposed RASSO (described in Section \ref {subsec:rasso}), which sharpens weak micro-Doppler responses before beamforming. All pipelines share identical MTI and static suppression, and the same downstream CA-CFAR configuration (Section \ref{subsec:ca-cfar}).

\subsection{Capon/MVDR Beamforming}
\label{subsec:capon}

After range--Doppler (RD) pre–processing and moving–target indication (MTI) (Section ~\ref{subsec:preproc}), each frame provides, for every range bin \(r\), a complex RD slice per receive channel, denoted \(Y_m(r,d)\in\mathbb{C}\), where \(m\in\{1,\ldots,N_{\!Rx}\}\) indexes Rx channels and \(d\) indexes the MTI–filtered Doppler bins. Following the Infineon geometry used in our DBF baseline, we employ the L–shaped three–Rx layout and form a two–element azimuth subarray from the \(\text{Rx1}\) and \(\text{Rx3}\) arms to estimate azimuth only to be consistent with our SDK workflow \cite{infineon_an141319, infineonAN155322}. For a fixed range bin $r$, we collect $N_D$ neighbouring Doppler indices around the MTI passband into a slow-time  \emph{snapshot matrix} (one column per Doppler cell, one row per Rx channel):

\begin{equation}
  X_r \;=\;
  \begin{bmatrix}
    Y_1(r,D_1) & \cdots & Y_1(r,D_{N_D})\\[-1mm]
    \vdots     & \ddots & \vdots\\[-1mm]
    Y_{N_{\!Rx}}(r,D_1) & \cdots & Y_{N_{\!Rx}}(r,D_{N_D})
  \end{bmatrix}
  \;\in\; \mathbb{C}^{N_{\!Rx}\times N_D}.
  \label{eq:snapshot}
\end{equation}

The Capon algorithm or minimum variance distortionless response (MVDR) beamformer is applied to the range-Doppler data, enabling the generation of precise range-Azimuth (RA) and range-Elevation (RE) feature maps \cite{10012054}. The algorithm estimates the Angle of Arrival (AoA) by solving an optimization problem that minimizes the output power $P_{\text{out}}$ while ensuring a distortionless response in the
desired target direction: 
\begin{equation}
\label{eq:mvdr-opt}
\min_{\mathbf w} \; \Bigg(  P_{\text{out}} = \frac{1}{2} \mathbf w^{\mathrm H}\mathbf R_r\,\mathbf w \Bigg)
\quad \text{s.t.} \quad
\mathbf w^{\mathrm H}\mathbf a(\theta)=1 .
\end{equation}
where $R_r$ is the sample spatial covariance at range $r$ that can estimated as following \eqref{eq:snapshot}:

\begin{equation}
  R_r \;=\; \frac{1}{N_D}\, X_r X_r^{\!H} \;\in\; \mathbb{C}^{N_{\!Rx}\times N_{\!Rx}}.
  \label{eq1:cov}
\end{equation}

and form its Moore–Penrose pseudoinverse $R_r^{-1}$ (no diagonal loading, $\delta=0$).
With azimuth steering vector:
\begin{equation}
  a(\theta) \;=\; \begin{bmatrix}1 \\ e^{-j\pi\sin\theta}\end{bmatrix},
  \label{eq:steer2el}
\end{equation}
the Capon power spectrum at range \(r\) and azimuth \(\theta\) is computed as
\begin{align}
  P_{\mathrm{Cap}}(r,\theta) \;&=\; \frac{1}{a(\theta)^{\!H}\,\widehat R_r^{-1}\,a(\theta)}.
  \label{eq1:capon}
\end{align}
which are similar to our previous work on in-vehicle monitoring \cite{Abedi2020OnTU} and fall detection \cite{11031826}. The only difference in ~\eqref{eq1:capon}'s notation is that we aggregate over all Doppler bins at that range to estimate $R_r$. 
We define the range–azimuth (RA) map as $\text{RA}_{\text{Capon}}(r,\theta)=|P_{\text{Capon}}(r,\theta)|$,
optionally smoothed over 5 frames and normalized to $[0,1]$ for visualization and CFAR. In the remainder of the paper, we refer to this Capon-based RA estimation ~\eqref{eq:mvdr-opt}–\eqref{eq1:capon} as our \textbf{Baseline method} or \textbf{Baseline RA} (\textbf{Conv RA}). When we include this method as the full signal-processing chain used for baseline results and ablations, we call the end-to-end chain the \textbf{Baseline pipeline} as shown in the Fig.~\ref{fig:flow_chart}.

\subsection{Proposed RASSO Algorithm}
\label{subsec:rasso}
 Semi-/quasi-static human motions employ subtle postural changes, and micro-motions concentrate Doppler energy in the tight neighbourhood of zero frequency. With a uniform slow-time FFT, the Doppler grid uses the same resolution everywhere, resulting in the bin near $f \approx 0$ being too coarse for the dynamics we are interested in. We propose RASSO, an optional wrapper that applies an invertible Doppler-domain non-linear warping and resampling of the Doppler axis before downstream spatial processing (Capon) and CFAR. Concretely, RASSO re-parameterizes each Doppler slice via the mapping $f \rightarrow D$ in ~\eqref{eq:gmap}-\eqref{eq:ginv} and re-samples the complex RD data on a uniform grid in  D as in ~\eqref{eq:realInt}-\eqref{eq:resample}, thereby allocating more samples around near-zero Doppler while preserving per-antenna phase. 
 
\begin{figure}[htb] 
\centering 
\includegraphics[width=\linewidth]{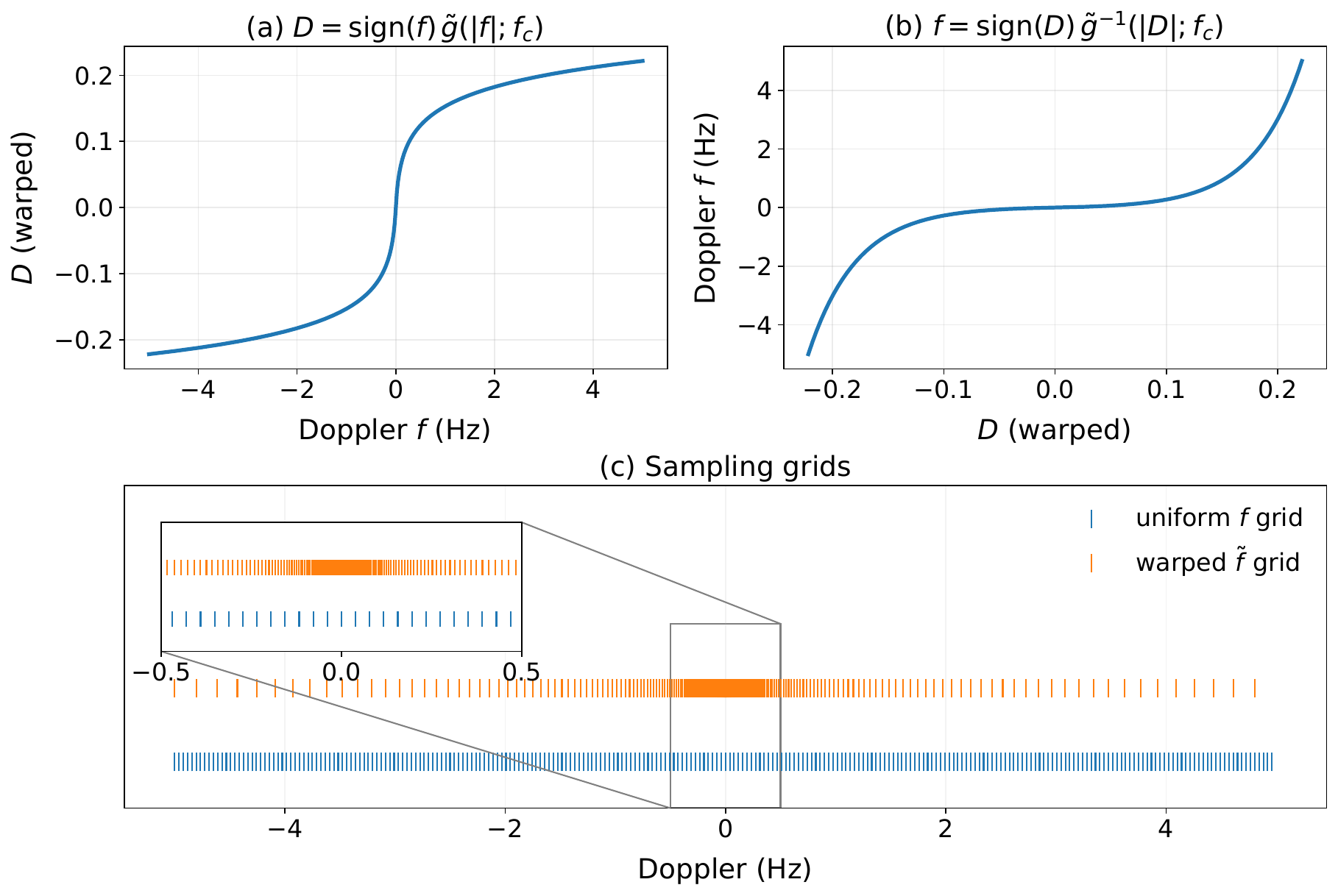}
\caption{Top subplots(left/right) depict forward and inverse mapping functions corresponding to ~\eqref{eq:gmap}–\eqref{eq:ginv}. The bottom panel shows how uniform spacing in D produces a non-uniform grid in 
$f$ that is densest around $f=0$ (zoom-in)}
\label{fig:rasso_mapping_frame_00100}
\end{figure}

\paragraph{Build the mapping grids} 
Log-like warps are the standard way to densify near-zero and sparsify far frequency sampling. 
Compared with generic time-frequency re-allocations, we maintain a monotone, invertible mapping, ensuring a one-to-one representation and preserving phase relationships used by subsequent beamforming. We densify sampling in $f \leq f_e$ (where the quasi-static energy is distributed), where $f_e > 0$ is the edge frequency, and compress the outer band, which is similar to the principle behind perceptual warping mel/log scales \cite{995829}.
For even symmetry in magnitude, we add the sign back so that directions are preserved, and we scale by $\frac{1}{log2}$ to normalize the mapping. In addition, the mapping must be DC-safe, meaning there is no singularity at $f = 0$. 
Let $f\in[-F_{\max},F_{\max}]$ denote the Doppler frequency of a given \((r,m)\) slice (after FFT shift) on the uniform grid, and let $D$ be the warped coordinate. We define this mapping 
and its inverse as:

\begin{align}
  D &= \operatorname{sgn}(f) \tilde{g}(f, f_e) = \operatorname{sgn}(f)\;
                     \frac{f_e}{\log 2}\,
                     \log\!\Big(1+\frac{|f|}{f_e}\Big),
                     \label{eq:gmap}\\
  f &= \operatorname{sgn}(f) \tilde{g}^{-1}(D, f_e) = \operatorname{sgn}(D)\;
                    f_e\!\left(2^{|D|/f_e}-1\right),
                        \label{eq:ginv}
\end{align}
Based on our empirical study, we tune the edge frequency and find that $f_e = 0.03$ is qualitatively sufficient for clear Range-Angle feature maps in almost all cases on our pilot dataset (with smaller $f$ values not improving further). Since the mapping is one-to-one, this prevents ambiguity in the conversion process between a continuous system and its discrete representations. In other word, the discrete–continuous–discrete step is stable and does not yield a significant amplification of errors in the discrete domain.
Fig.~\ref{fig:rasso_mapping_frame_00100} (top left and right) visualizes \eqref{eq:gmap} and \eqref{eq:ginv} as the forward warp and the inverse used for resampling, respectively. The grid plot in Fig.~\ref{fig:rasso_mapping_frame_00100} (bottom) shows how uniform steps in $D$ (in blue) produce a non-uniform grid (in orange) in $f$, that is much denser near zero (see zoom insert).

\paragraph{Complex resampling of RD}
For each range bin $r$ and antenna $m$, we denote $Y_m[r,f]$ as MTI-filtered Doppler slice on the uniform FFT grid $\{f_l\}_{l=0}^{N_D-1}$ (FFT-shifted). We resample it into uniform grid $D \in \{-N_d/2,...,N_d/2-1 \}$ using a inverse map (\ref{eq:ginv}). We implement vectorized complex linear interpolation on the Doppler axis by interpolating the real and imaginary parts separately:
\begin{equation}
    \mathcal{R} \{\tilde{Y}_m[r,D_k]\} \approx \mathrm{Interp1D}(\{f_l\}, \mathcal{R}\{Y_m[r,f_l]\}, g^{-1}(D_k;f_e)) 
    \label{eq:realInt}
\end{equation}
\begin{equation}
    \mathcal{I} \{\tilde{Y}_m[r,D_k]\} \approx \mathrm{Interp1D}(\{f_l\}, \mathcal{I}\{Y_m[r,f_l]\}, g^{-1}(D_k;f_e)) 
    \label{eq:imgInt}
\end{equation}
and then we compute $\tilde{Y}_m[r,D_k] = \mathcal{R}\{\cdot\} + j\mathcal{I\{\cdot\}}$ with zero-pads outside $[f_0,f_{N_D-1}]$. Since vectorized real/imaginary interpolation and Capon beamforming remain unchanged, the warp's complexity is $O(RN_DN_{RX})$ linear-time per frame. In practice, wall-time overhead is dominated by the Capon covariance inversion. Finally, stacking over m and r yields the RASSO-enhanced RDC $\tilde{Y[r,D_k,m]}$.  

\begin{equation}
  \widetilde{Y}_m[r,D] \;\approx\; Y_m\!\big[r,\; g^{-1}(D;f_e)\big].
  \label{eq:resample}
\end{equation}

\begin{figure}[htb] 
\centering 
\includegraphics[width=\linewidth]{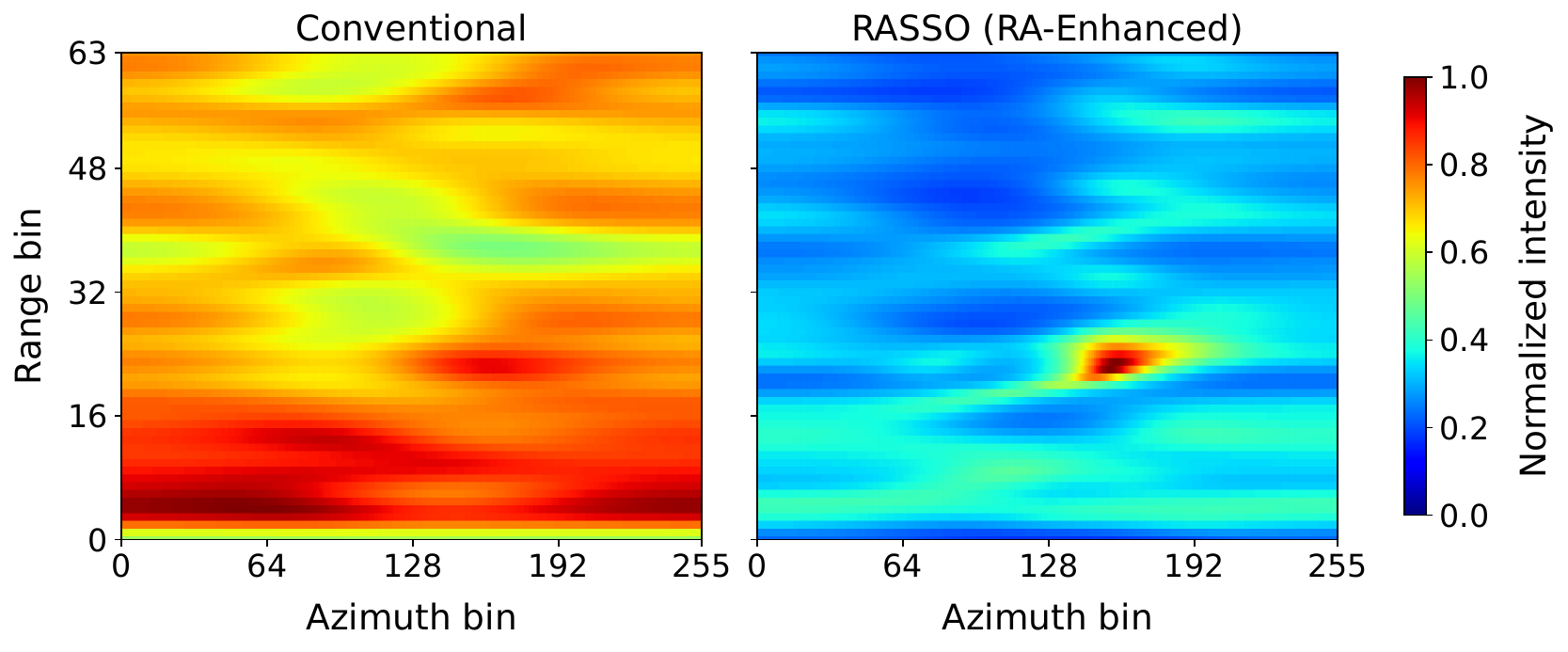}
\caption{Range-Azimuth feature map (frame 100) of subject three lying on the floor using Conventional (left) and RASSO enabled (right) methods.}
\label{fig:rasso_effect_frame_00100}
\end{figure}

In other words, RASSO replaces each Doppler slice $Y_m[r,f]$ defined on the uniform FFT grid $\{f_\ell\}$ with $\tilde{Y}_m[r,D_k] \approx Y_m\bigl[r, g^{-1}(D_k; f_e)\bigr], $ i.e., an invertible non-linear change of coordinates on the Doppler axis, implemented via complex-valued interpolation, without modifying the underlying
complex field. Fig.~\ref{fig:rasso_effect_frame_00100} compares Baseline versus RASSO-Enhanced Range-Angle feature maps (frame 100) of Subject 3 lying on the floor. It can be seen that after applying RASSO to the diffused Conventional map, the same scene exhibits a compact, blob-like target with suppressed low-angle smear, consistent with more Doppler samples being devoted to quasi-static before covariance estimation and beamforming. 

\subsection{Cell-Averaging CFAR (CA–CFAR) on RD/RA Maps}
\label{subsec:ca-cfar}
\begin{figure}[htb]
\centerline{\includegraphics[width=\linewidth]{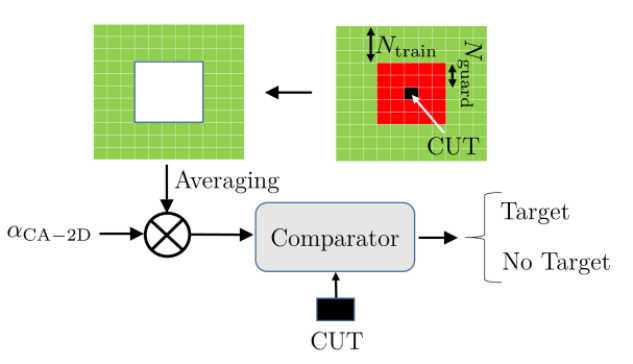}}
\caption{2D CA-CFAR \cite{10289281}.}
\label{fig:2D-CFAR-plot}
\end{figure}

Let $Y(i,j)$ denote a generic 2-D radar feature map - either a Range-Doppler ($Y(r,f_D)$) or Range-Azimuth ($Y(r,\theta)$) case, we perform detection on these image-like feature maps using the \emph{cell–averaging CFAR} (CA–CFAR) as described in \cite[section 7.2-7.3]{Miller2009FundamentalsOR}. For each cell under test (CUT),  
we place a square window of side $2H+1$ cells centred at CUT. A \emph{guard band} of $G$ cells per side protects the CUT and its intermediate neighbours from leaking target energy into the noise estimate. The \emph{training ring} therefore had width $T$ cells per side, so that the training region is the outer ring of the $(2H+1)\times(2H+1)$ window after removing the $(2G+1)\times(2G+1)$ guard as visualized in the Fig.~\ref{fig:2D-CFAR-plot}. Let $\mathcal{T}(r,c)$ be the set of training cells, the sample average estimates the local mean interference power
\begin{equation}
\widehat{\mu}(r,c) = \frac{1}{N_\text{train}} \sum_{(i,j)\in\mathcal{T}(r,c)}Y(i,j)
\end{equation}
\begin{equation}
N_\text{train} = (2H{+}1)^2 - (2G{+}1)^2
\end{equation}
A detection is declared at $(r,c)$ when
\begin{equation}
  Y(r,c) \;>\; \Gamma(r,c), \qquad \Gamma(r,c) \triangleq k\,\widehat{\mu}(r,c),
  \label{eq:cafar-thresh}
\end{equation}
where $k>0$ is a sensitivity factor  (or $\alpha$ in Richards \cite[Eq.(7.11)]{Miller2009FundamentalsOR} that controls the effective false-alarm rate. In preliminary sweeps experiment, we varied $G\in\{2,4,6\}$ and $T\in\{6,10,14\}$, and found performances stable once $ T\ge 10$. 
Thus, unless specified in Section ~\ref{sec:result}, we  vary $k$ use $G{=}4$, $T{=}10$, which yields $H{=}14$ and $N_\text{train}{=}760$ training cells per CUT.

\subsection{From Pixel Detections to Frame-Level Presence}
\label{sec:frame-presence}
\begin{figure*}[!t] 
  \centering    \includegraphics[
    width=\textwidth,
    height=.31\textheight,   
    keepaspectratio,         
    trim=0 6pt 0 6pt, clip   
  ]{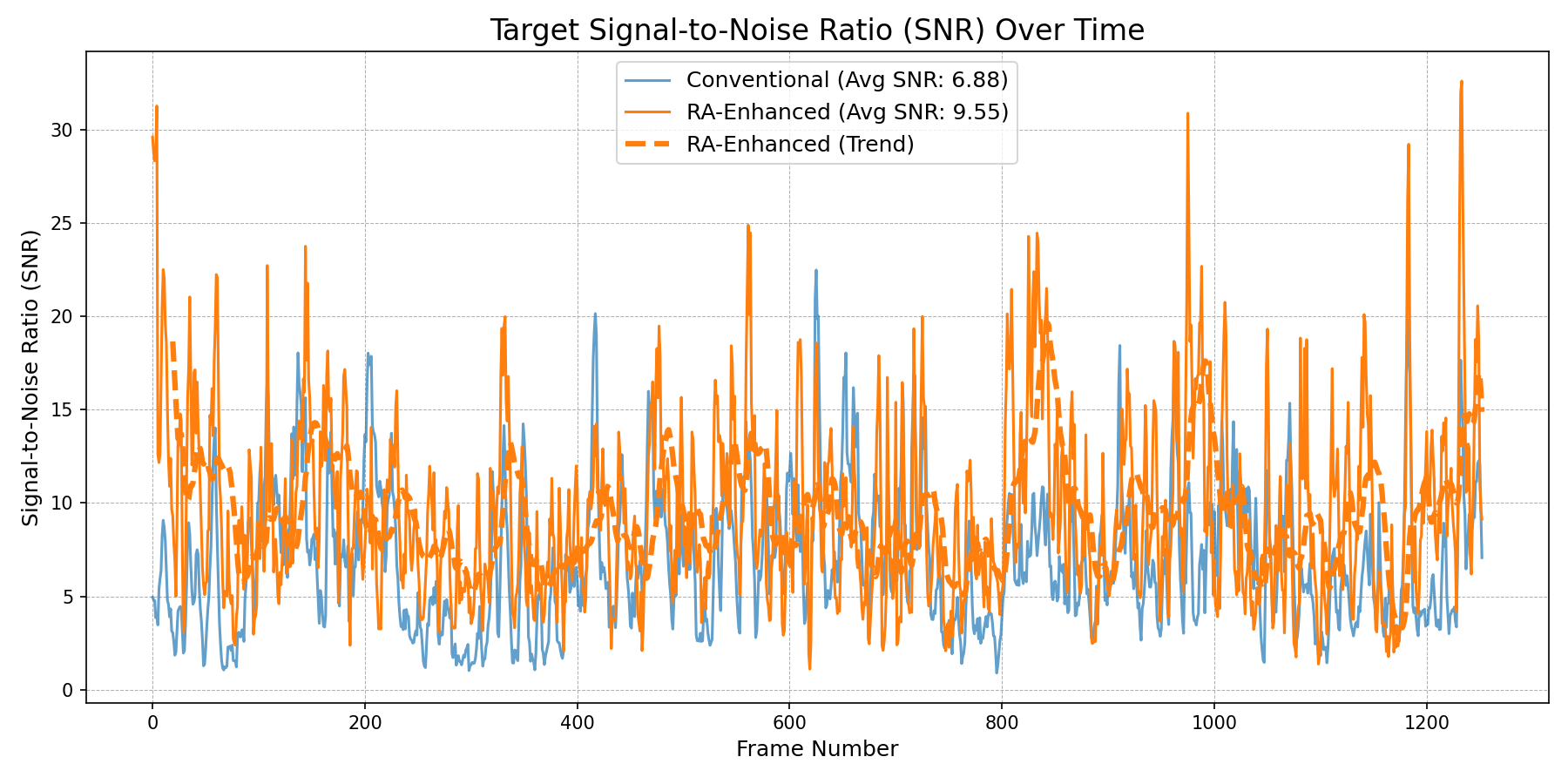}
  \caption{Signal-to-noise ratio plot corresponding to subject lying on the floor in Fig.~\ref{fig:result_1253_frames}.}
  \label{fig:SNR_result}
\end{figure*}
\begin{figure*}[!t] 
  \centering    \includegraphics[
    width=\textwidth,
    height=.6\textheight,   
    keepaspectratio,         
    trim=0 6pt 0 6pt, clip   
  ]{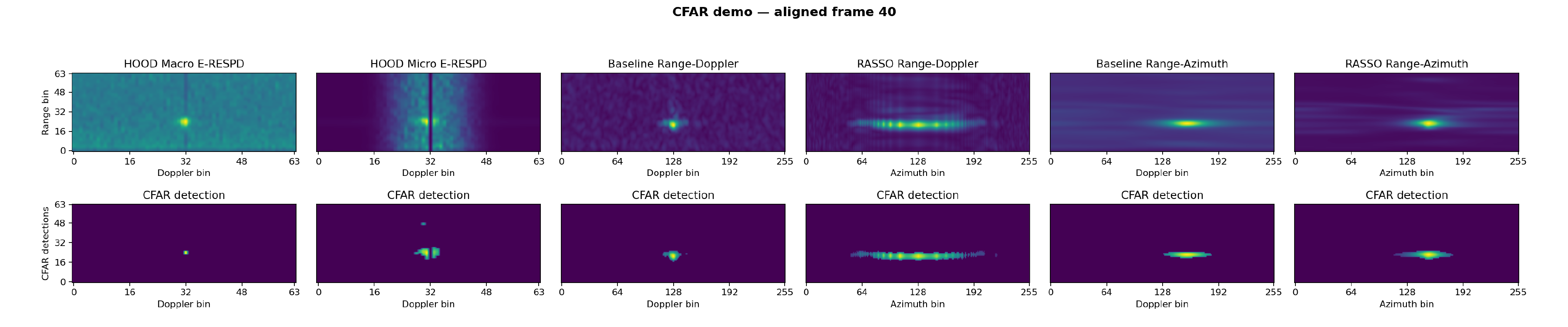}
  \caption{Visualization of 6 feature maps on first row and their corresponding CFAR detections on second row (from left to right): Macro E-RESPD, Micro E-RESPD, Baseline Range Doppler, RASSO-enhanced Range-Doppler, Baseline Range-Azimuth, RASSO-enhanced Range-Azimuth.}
  \label{fig:compare_all_frame_40}
\end{figure*}

In general, our end-task is a frame-level presence flag (person present vs. empty), which is a system-level quantity not predicted by any single CUT.
We implement integral-image realization from \eqref{eq:cafar-thresh} with $O(1)$ computation per CUT. This would be identical across the methods described in Section \ref{sec:result}. We adapt a simple, uniform image-processing post-step to convert the CFAR mask into a stable binary frame decision, as described in \cite[Section 7.7.5]{Miller2009FundamentalsOR}. Following the CA-CFAR test at each pixel, we apply a single $3{\times}3$ morphological opening to suppress isolated speckle. Let $I_c$ denote the post-opening mask on the map (RD or RA). We label the frame as positive if the mask contains at least one connected component with area $\geq A_{min}$ pixels; otherwise, it is negative as follows: 
\begin{equation}
D_{\text{presence}}=
\begin{cases}
1, & \displaystyle \text{if } \max_{\mathcal{C}\in \mathrm{CC}(I)} |\mathcal{C}| \;\ge\; A_{\min},\\[6pt]
0, & \text{otherwise.}
\end{cases}
\tag{27}
\end{equation}
where $CC(I)$ lists all 8-connected components in the mask and $|\mathcal{C}|$ is the component area (in pixels). We use $A_{min} =
12$ in our experiment setting. We set the area threshold to be just larger than the biggest speckle patch that survives CA-CFAR and a 3×3 opening, yet smaller than the smallest true target blob we observe across all six upstream maps we compare (E-RESPD Macro/Micro E-RESPD, Conventional RD/RA, and RASSO-enhanced RD/RA) as shown in Fig.~\ref{fig:compare_all_frame_40}. This is agnostic to which feature map is used upstream (DBF, Capon, or RASSO), and converts a large set of per-pixel tests with a nominal per-CUT false-alarm rate into a single, robust frame decision.

\begin{figure*}[!t] 
\label{fig:FAR_DBF}
  \centering
  \includegraphics[
    width=\textwidth,
    keepaspectratio,          
    trim={0 0 0 0pt},        
    clip                      
  ]{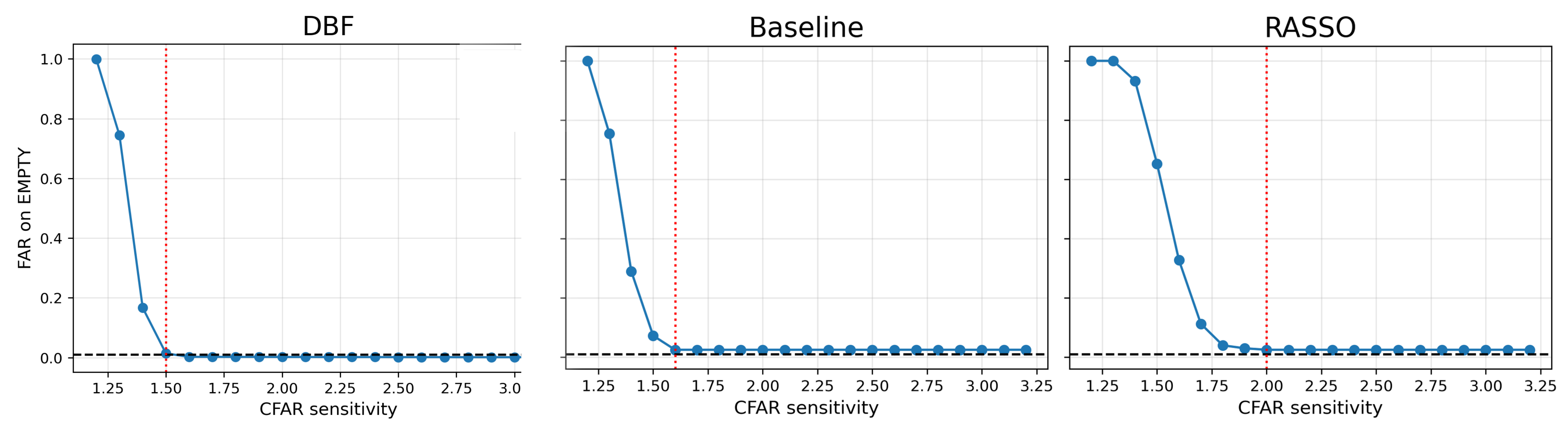}
  \caption{False–alarm rate (FAR) on empty-room sequences as a function of the CA–CFAR sensitivity \(k\) for the three processors: (\textbf{left}) SDK Digital Beamforming (DBF), (\textbf{middle}) our Conventional baseline, and (\textbf{right}) our RASSO-enhanced pipeline. Curves are computed on negatives only (no person present), after MTI (\(\alpha{=}0.01\)), five-frame mean smoothing, \([0,1]\) normalization, and CA–CFAR with guard \(G{=}4\) and train \(T{=}10\). Dots indicate individual sequence measurements; the solid line is the mean FAR. The red vertical dashed line marks the operating point chosen for each method (smallest \(k\) at which mean FAR meets the target constraint used throughout the paper).}
  \label{fig:far_sweep_triptych}
\end{figure*}

\section{RESULTS}
\label{sec:result}
In this section, we present the qualitative study by first visualizing feature maps from 2-minute recording data before and after applying RASSO. In addition, we quantitatively compute the signal-to-noise ratio (SNR) corresponding to this.
To demonstrate further the efficacy of RASSO, we study ablation of RASSO on our baseline preprocessing method and compare against SDK DBF from Infineon vendor (Section \ref{subsec:dbf}) and Micro/Macro-RESPD method \cite{kahya2023mcroodmulticlassradaroutofdistribution,10789192}. We also examine the effectiveness of applying RASSO to cross-subject data using knowledge-driven and data-driven approaches. Finally, we conduct uncertainty quantification of our best method to demonstrate the superior effect of RASSO.

\subsection{Visualization and SNR analysis of RASSO method}
\begin{figure}[htb]
\centerline{\includegraphics[width=\linewidth]{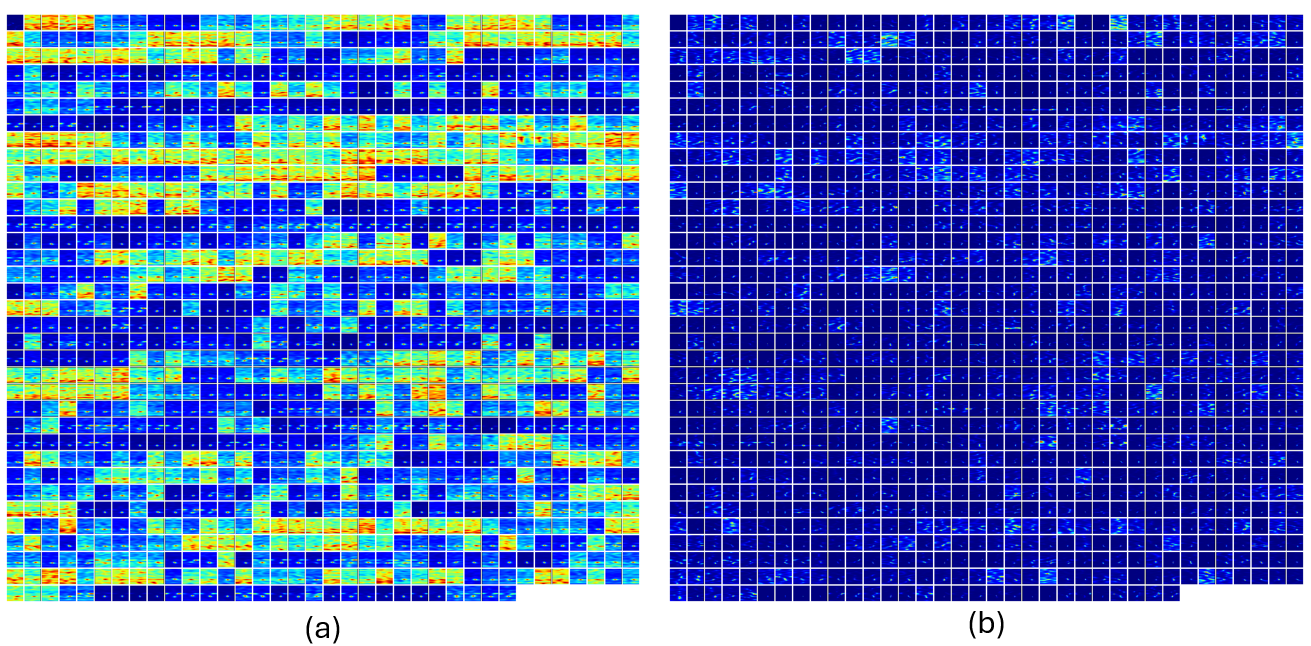}}
\caption{Range-Azimuth feature maps of 2-minute data collected when the subject was lying on the floor (a) before applying RASSO and (b) after applying RASSO. The RASSO-processed maps in (b) exhibit noticeably reduced background clutter and enhanced target energy, indicating an improved signal-to-noise ratio that supports more robust recognition.}
\label{fig:result_1253_frames}
\end{figure}

We first visualize the Range-Azimuth feature maps from a 2-minute data collection for 1 participant in the Lay On Floor case. As shown in Fig.~\ref{fig:result_1253_frames}, few conventional RA feature maps are burdened in the background and dominated by clutter (in yellow/cyan). At the same time, RASSO effectively enhances the signal and improves the overall robustness of the signal.

To quantify performance over time, we compute a framewise SNR. For each detected bounding region $\mathcal{B}$ and a local background ring $\mathcal{N}$ (the CFAR training set excluding $\mathcal{B}$), we define
\begin{equation}
\mathrm{SNR} = \frac{\overline{Y}_{\mathcal{B}} - \overline{Y}_{\mathcal{N}}}{\sigma_{\mathcal{N}}},
\label{eq:snr}
\end{equation}
a z-score robust to slow gain drifts. As shown in Fig.~\ref{fig:SNR_result}, the RASSO curve is consistently higher and smoother than the conventional curve, indicating stronger signal concentration and reduced background variance. To quantify improvement, we computed the targeted Signal-to-Noise Ratio (SNR) by taking the ratio of mean power within the detected bounding box before and after applying RASSO, and the results are demonstrated in Fig.~\ref{fig:SNR_result}. 
The RASSO-enhanced SNR (orange) is consistently higher than conventional SNR (blue) with average values of 6.88 and 9.55 dB, respectively. 

\subsection{Digtial Beam Forming comparison}\label{sec:dbf_comp}
We compare our baseline pipeline with and without RASSO against Digital Beamforming (DBF) from Infireon's Software Development Kit (SDK). All processors share the same pre- and post-processing. To keep up with the compatibility with our conventional pipeline used in  \cite{10784889} and compare to the Digital Beamforming pipeline provided (which also uses MTI), we use the same MTI as described in Section \ref{subsec:rdm_generation} with a very small $\alpha$ (0.01 is chosen) to balance between very conservative DC cluster removal as well as keep our quasi-static signal alive. For a fair comparison, each method’s CFAR sensitivity $s$ is tuned only on negative (empty-room) sequences and sweep $s$ on a grid and choose $s$  that minimizes $| \text{FAR}(s) - \text{FAR}_\text{target} |$ on the pooled empty-room set ($\text{FAR}_\text{target}$=0.01) for all three processors. Balanced Accuracy (BA) is then computed on the evaluation set as \(\text{BA}=\tfrac{1}{2}(\text{True Positive Rate (TPR)}+\text{True Negative Rate (TNR)})\).

\begin{figure}[htb]
\centerline{\includegraphics[width=\linewidth]{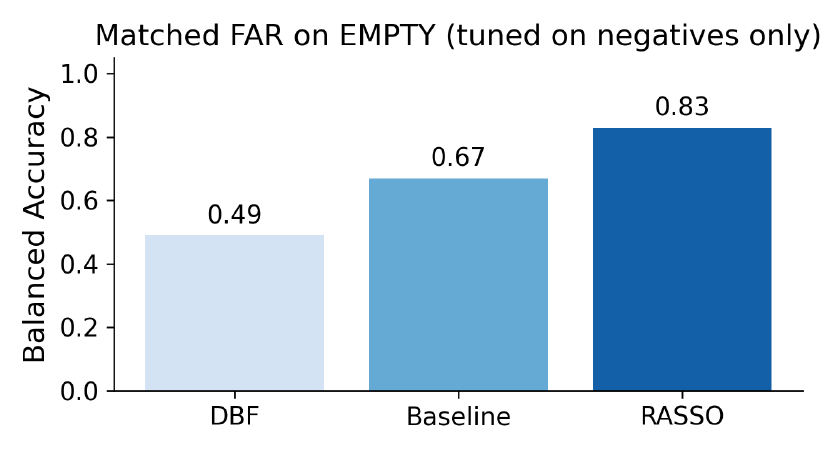}}
\caption{Balanced Accuracy (BA) at matched-FAR operating points for SDK DBF, our Conventional baseline, and our RASSO-enhanced processor. For fairness, the CA–CFAR sensitivity \(k\) for each method is tuned on empty-room sequences only to meet the same FAR target as Fig.~\ref{fig:far_sweep_triptych}.
}
\label{fig:DBF_comp_bar}
\end{figure}

\begin{figure}[t]
\centerline{\includegraphics[width=\linewidth]{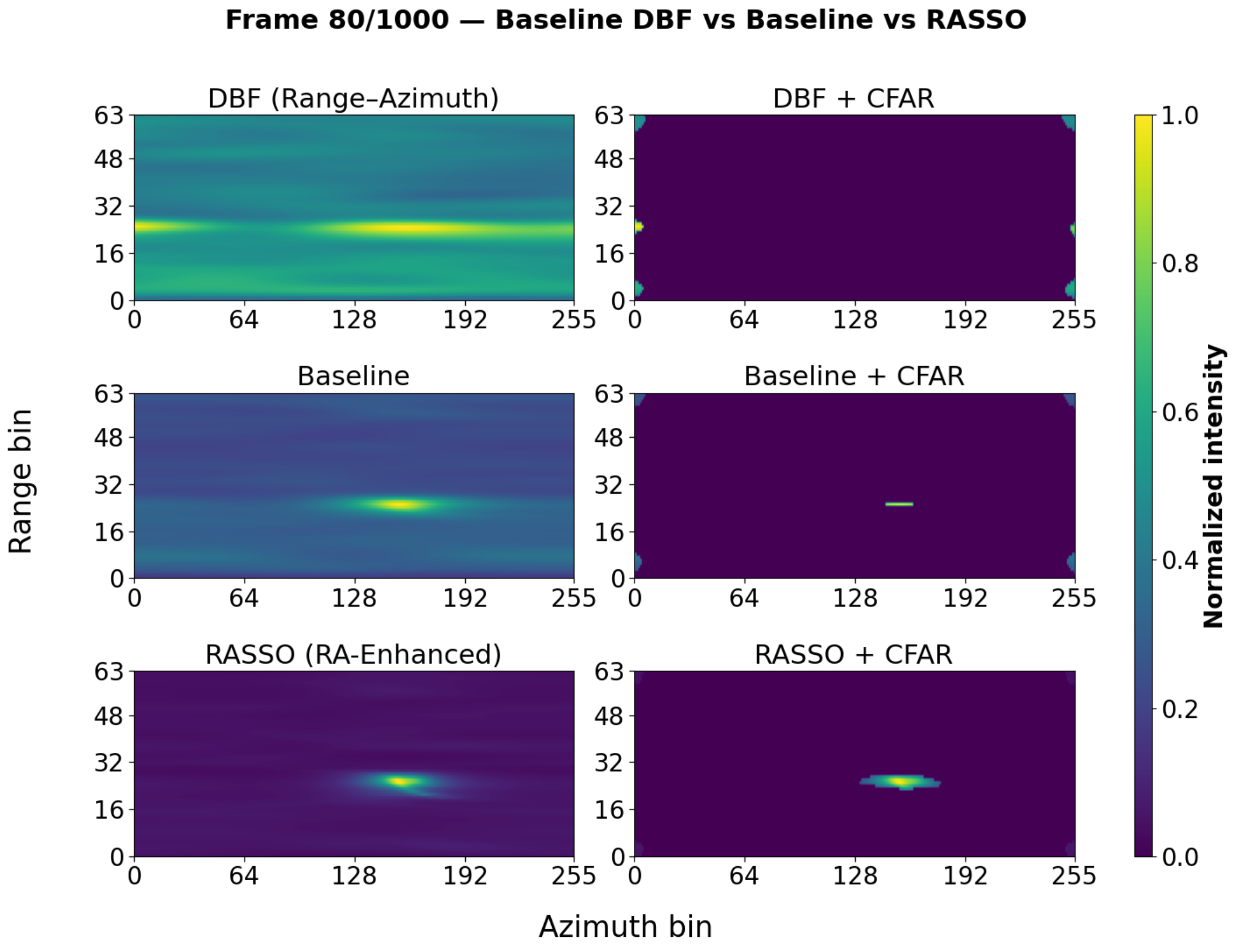}}
\caption{(\textbf{Left Column}) Range-azimuth feature map (frame 80) of the subject lying on the floor obtained by (\textbf{Top}) Digital Beam Forming (DBF), (\textbf{Middle}) Our Conventional, (\textbf{Bottom}) RASSO enhanced on Our Conventional. The right column show their corresponding CFAR detection.}
\label{fig:frame_80}
\end{figure}
To make qualitative comparisons meaningful, we render all range–azimuth (RA) maps using a standard colour scale after applying mean smoothing overd a 5-frame window and $[0,1]$ normalization. This is the same normalization used in prior work (e.g., Hajar et al. \cite{Abedi2020OnTU}).
Representative frames are shown in Fig. \ref{fig:frame_80} (left column). In the DBF baseline, the energy blob associated with the subject typically spans a broader set of azimuth bins, indicating a less concentrated angular response. Our baseline processor, however, yields a visibly tighter lobe, and RASSO further concentrates the energy, reducing the angular spread and sidelobe floor. These effects are consistent across activities we inspected and across rooms.
Using two empty-room files and very small pilot activity files for demonstrating proof-of-concept, the sensitivity values selected by the above protocol were within the ranges illustrated in Fig. \ref{fig:frame_80} (right column) and Fig. \ref{fig:DBF_comp_bar}. We evaluated these matched-FAR operating points on our minimal pilot dataset for proof-of-concept and reported Balanced Accuracy rather than raw accuracy. Balanced Accuracy (BAcc) is the average of the true-positive rate (occupied) and the true-negative rate (empty).  RASSO achieves the highest BAcc, followed by our baseline processor, and the SDK DBF's accuracy is the lowest.
It is consistent with the FAR-sensitivity sweeps (Fig. \ref{fig:far_sweep_triptych}), where RASSO reaches the low-FAR regime at milder sensitivity than DBF.

\subsection{Benchmarking of RASSO against E-RESPD}
Fixing a single CA-CFAR threshold $k$ might not be a fair way to compare different processing chains: each feature map (HOOD E-RESPD Macro/Micro, our Conventional RD/RA, and RASSO-enhanced RA) has different normalization, dynamic-range compression, and residual clutter statistic. In addition, the same $k$  does not imply the same false-alarm probability $\overline{P}_\text{FA}$ \cite[Sec.~7.3]{Miller2009FundamentalsOR}. Specifically, we sweep $k$ on a dense grid and trace recall versus FAR curves on a per-frame basis.  The area under this curve (AUC) plays the same role as AUROC in the MCROOD/HOOD papers \cite{kahya2023mcroodmulticlassradaroutofdistribution,10789192}. For each $k$ we count over all frames:
\begin{itemize}
  \item \textbf{Recall (TPR):} fraction of human presence file frames with $\geq$\,1 valid cluster;
  \item \textbf{False Alarm Rate} (\textbf{FAR}): fraction of empty room files frames with $\geq$\,1 valid cluster.
\end{itemize}

\begin{figure}[htb]
\centerline{\includegraphics[width=\linewidth, height=\textheight, keepaspectratio]{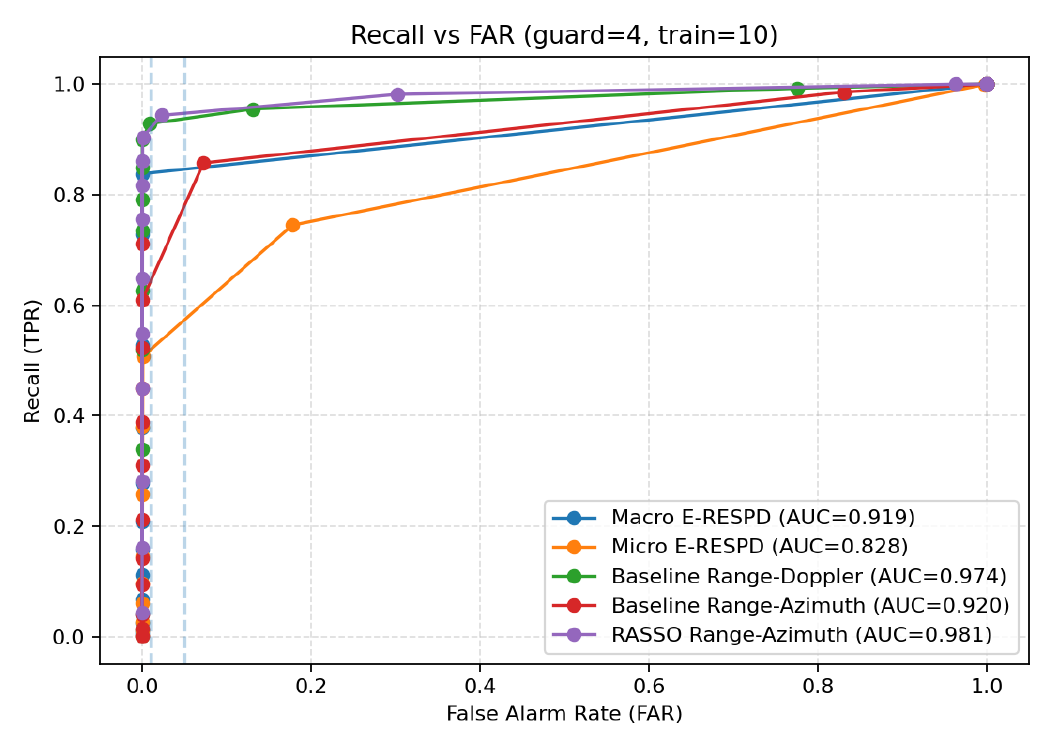}}
\caption{Recall–FAR characteristics obtained by sweeping the CA–CFAR sensitivity \(k\) for five feature maps: HOOD Macro E-RESPD, HOOD Micro E-RESPD, Baseline Range–Doppler, Baseline Range–Azimuth, and RASSO Range–Azimuth. The legend reports the area under the Recall–FAR curve (AUC). Light blue vertical dashed lines indicate reference operating regions at FAR \(=1\%\) and \(5\%\) used for pointwise comparisons.}
\label{fig:recall_far_curves_all}
\end{figure}

Fig.~\ref{fig:recall_far_curves_all} compares and visualizes ROC-like lines and trapezoidal area for the above 6 methods.  and we also report recall at practical operating points (FAR = 1\% and 5\%) as in Table \ref{tab:recall_far_table}. Although RASSO is not originally designed for boosting the Range-Doppler map, it is interesting to include it for study reference purposes. Using this evaluation, our RASSO-RA front-end enhances the Range-Angle feature map method and overtakes the HOOD baselines and our conventional variants. It achieves the highest AUC (\textbf{0.981}) and highest Recall at both FAR targets ($\textbf{0.920}$ @1\%, $\textbf{0.947}$ @5\%). This indicates that warping near-zero Doppler to reduce DC before spatial processing (RASSO) followed by adaptive MVDR beamforming, improves separability at low FAR.

\begin{table}[H]
\centering
\caption{Summary of AUC of the Recall–FAR curve and Recall at two fixed false-alarm operating points (FAR \(=1\%\) and \(5\%\)) among 5 methods. Best values per column are shown in \textbf{bold}.}
\label{tab:recall_far_table}
\resizebox{\linewidth}{!}{
\begin{tabular}{lccc}
\toprule
Method & AUC & Recall@FAR=1\% & Recall@FAR=5\%\\ 
\midrule
Macro E-RESPD & 0.919 & 0.839 & 0.845\\
Micro E-RESPD & 0.828 & 0.520 & 0.573 \\
Range-Doppler & 0.974 & 0.829 & 0.837 \\
Baseline Range-Azimuth & 0.920 & 0.643 & 0.780 \\
\midrule
RASSO Range-Azimuth & \textbf{0.981} & \textbf{0.920} & \textbf{0.947}\\
\bottomrule
\end{tabular}}
\end{table}

Our baseline RD also yields AUC $0.974$ with high Recall at 1–5\% FAR. The Macro and Micro E-RESPD achieve AUCs of $0.919$ and $0.828$, respectively. These curves motivate using RASSO-RA as the feature map for the remaining of the paper.

\subsection{Cross-subject Evaluation}
While a $k$ sweep is suitable for benchmarking, deployment requires sticking to an operating point. We therefore conduct a RASSO ablation study, sweep $k$ on a development subject (Subject 1) and evaluate that fixed $k$ on unseen subjects (Subjects 2–4) recorded in the same living room from different vantage points. We select $k$ to maximize macro-F1 (unweighted mean of class F1 for empty room and 1 person) to balance the misses and false alarm rate without privileging one class. Similar to our DBF comparison study in the Section \ref{sec:dbf_comp}, our baseline pipeline induces higher sensitivity $k = 1.6$ with RASSO and $k=1.2$ without RASSO. The results are demonstrated in Table \ref{tab:CA-CFAR-result}.

\begin{table}[h!]
\centering
\caption{CA-CFAR Evaluation Across Subjects\\($k=1.2$ for Baseline Range-Azimuth and $k=1.6$ for RASSO-Enhanced).  Best values for each F1 metrics per subject are shown in \textbf{bold}.}
\label{tab:CA-CFAR-result}
\resizebox{\linewidth}{!}{%
\begin{tabular}{@{} l|l c c c @{}}
\toprule
\textbf{Subject} & \textbf{Variant} & \textbf{F1 Macro} & \textbf{F1 (empty)} & \textbf{F1 (1-person)} \\
\midrule

\multirow{2}{*}{\makecell[l]{Subject 1\\(Development Set)}} 
 & Baseline RA & 0.843 & 0.771 & 0.915 \\
 & \textbf{RASSO RA} & \textbf{0.932} & \textbf{0.897} & \textbf{0.967} \\
\midrule

\multirow{2}{*}{\makecell[l]{Subject 2\\(Generalization Test Set)}} 
 & Baseline RA & 0.811  & 0.653 & 0.970 \\
 & \textbf{RASSO RA} & \textbf{0.924} & \textbf{0.857}  & \textbf{0.990}  \\
\midrule

\multirow{2}{*}{\makecell[l]{Subject 3\\(Generalization Test Set)}} 
 & Baseline RA & 0.842 & 0.720 & 0.964 \\
& \textbf{RASSO RA} & \textbf{0.948} & \textbf{0.907} & \textbf{0.990}\\
 \midrule

 \multirow{2}{*}{\makecell[l]{Subject 4\\(Generalization Test Set)}} 
 & Baseline RA  & 0.801 & 0.652 & 0.950  \\
 & \textbf{RASSO RA} & \textbf{0.863} & \textbf{0.863} & \textbf{0.985} \\

\bottomrule
\end{tabular}%
}
\end{table}
It can be seen from the Table~\ref {tab:CA-CFAR-result} that RASSO-RA consistently lifts performance for every subject and for both classes, resulting in an improvement of F1 Macro of $\approx$ 6\% to 10\%. Crucially, $F1_{empty}$ increases markedly, confirming that RASSO suppresses quasi-static clutter and clarifies the spatial covariance used by CFAR. With RASSO-enabled, across subjects, 
$F1_{1-person}$ remains very high (typically $\geq 0.97$), but 
$F1_{empty}$ are slightly lower, yielding macro-F1 range of 0.85–0.9 for empty room class. In privacy-preserving monitoring for nursing homes and long-term-care facilities, both missed detections (false negatives) and false alarms incur high costs. Missed detections can delay assistance, excessive false alarms drive alarm fatigue and affect trust in the system. A deployable solution, therefore, needs simultaneously high recall and low FAR with real-time latency and generalizes well for different residents.

\begin{figure*}[!t] 
  \centering    \includegraphics[
    width=\textwidth,
    height=.5\textheight,   
    keepaspectratio,         
    trim=0 6pt 0 6pt, clip   
  ]{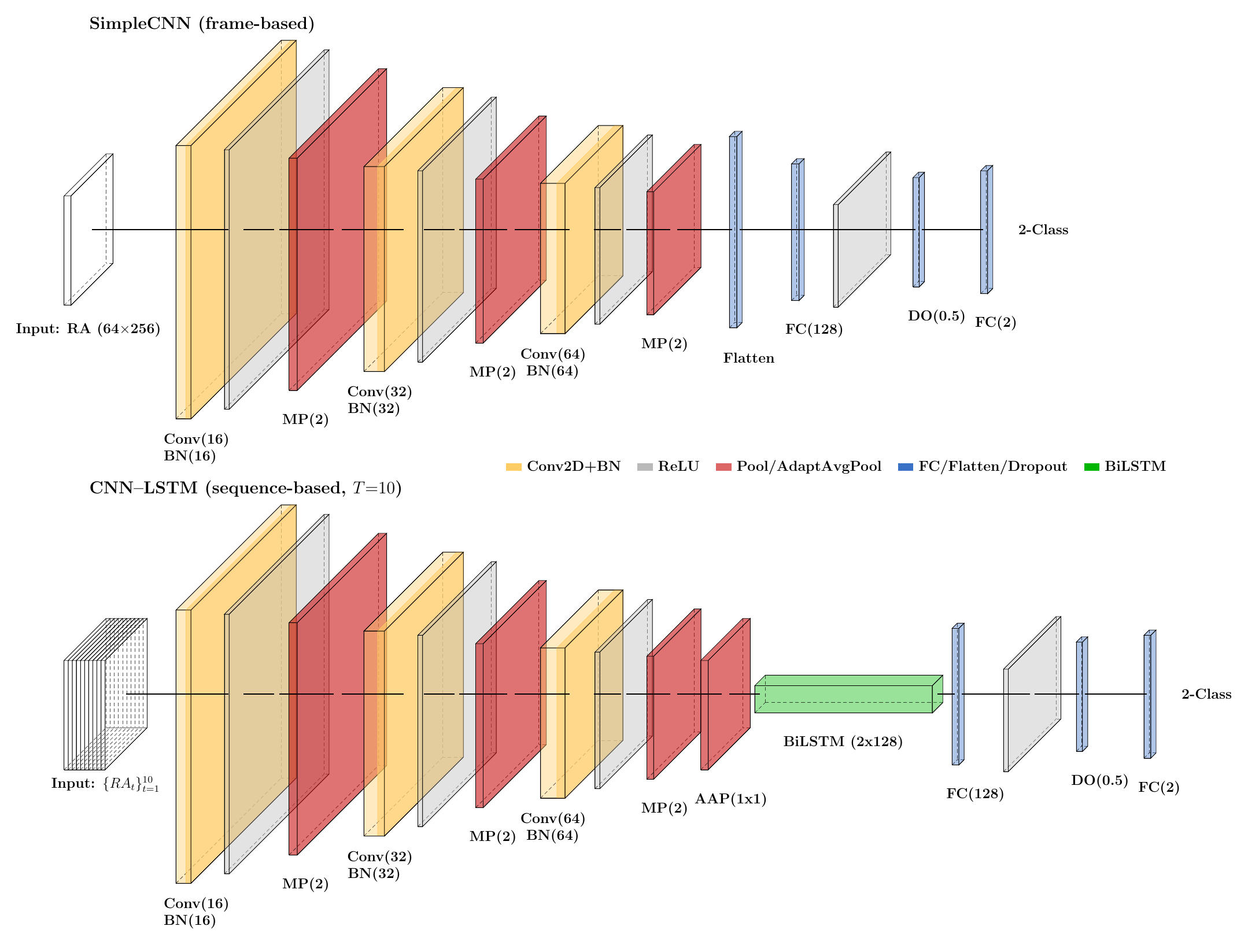}%
  \caption{Architecture of models (top): SimpleCNN and (bottom) CNN-LSTM.}
  \label{fig:model_archs}
\end{figure*}

\subsection{Data-Driven Classification}
\label{sub:Data-Driven-Classification}

To reduce these nuisance detections while retaining high sensitivity to human micro-motion, we complement the model-driven pipeline with lightweight, data-driven classifiers trained directly on the same RA maps. RASSO reshapes the Doppler axis and suppresses near-static clutter, producing range–Azimuth maps that are sparser with higher output SNR as shown in Fig.~\ref{fig:result_1253_frames}. These properties are ideal for learning: the discriminative signal is concentrated, while the nuisance structure is reduced. Thus, we design our own 2 models that reveal the impact of RASSO on learnability and generalization. 
\paragraph{SimpleCNN (frame-based) model}
The input is a single normalized RA heatmap ($64 \times 256)$. We first implemented a Convolution Neural Network (CNN) based model called SimpleCNN for each frame classification. Its architecture is visualized as Fig.~\ref{fig:model_archs} (top).  The model comprises a very shallow depth of three $3 \times 3$ convolution blocks with channel width $[16,32,64]$, each followed by batch normalization layers, ReLU activation functions and $2 \times 2$ max-pooling.  We trained SimpleCNN using Adam optimizer, early stopping on validation loss a 128-unit fully connected layer (ReLU, dropout $p=0.5$ to reduce overfitting) precedes 2-way softmax. For the training procedure, subject 1's dataset is split 80\,\%/10\,\%/10\,\% into training/validation/test for held-out evaluation, while all subjects 2-4 data are used for cross-subject generalization. The label space is binary 
${empty\_room,person\_present}$ and performance is reported as overall accuracy and per-class F1 (harmonic mean of precision and recall), which balances false alarms and misses. Table~\ref{tab:rasso_frame} summarizes the frame-based results. Across all subjects, RASSO-based RA maps consistently surpass the Baseline RA in terms of accuracy and F1, with the largest gains observed on the challenging cross-subject setting, where RASSO improves F1 (person-present) while maintaining high specificity on empty frames. These results indicate that the proposed Doppler warping not only sharpens classical CFAR-based detection but also provides more discriminative inputs for lightweight CNN classifiers without changing the network architecture or training protocol.
\begin{table}[h!]
\centering
\caption{RASSO Results – Frame Based (SimpleCNN)}
\label{tab:rasso_frame}
\resizebox{\linewidth}{!}{%
\begin{tabular}{@{} l|l c c c @{}}
\toprule
\textbf{Subject} & \textbf{Variant} & \textbf{Acc. (\%)} & \textbf{F1 (empty)} & \textbf{F1 (1-person)} \\
\midrule
\multirow{2}{*}{\makecell[l]{Subject 1\\(Held-out Test Set)}} 
 & Baseline RA  & 0.9100 & 0.9221 & 0.8934 \\
 & \textbf{RASSO RA} & \textbf{0.9536} & \textbf{0.9582} & \textbf{0.9252} \\
\midrule
\multirow{2}{*}{\makecell[l]{Subject 2\\(Generalization Test Set)}} 
 & Baseline RA & 0.9323 & 0.8509  & 0.9562\\
 & \textbf{RASSO RA} & \textbf{0.9687} & \textbf{0.9252} & \textbf{0.9802} \\
\midrule
\multirow{2}{*}{\makecell[l]{Subject 3\\(Generalization Test Set)}} 
 & Baseline RA  & 0.9250 & 0.8814 & 0.9451 \\
 & \textbf{RASSO RA} & \textbf{0.9678} & \textbf{0.9456} & \textbf{0.9771} \\
 \midrule
 \multirow{2}{*}{\makecell[l]{Subject 4\\(Generalization Test Set)}} 
 & Baseline RA & 0.8939 & 0.8394 & 0.9208 \\
 & \textbf{RASSO RA} & \textbf{0.9942} & \textbf{0.9899} & \textbf{0.9960} \\
\bottomrule
\end{tabular}%
}
\end{table}
\paragraph{CNN-LSTM (sequenced-based) model}
Semi-static occupancy often manifests as very low-Doppler micro-motions (postural sway, subtle limb readjustments, respiration) that unfold across multiple frames. Thus, classification based on individual frames may miss these cues. LSTMs are expressly designed to learn long/short-term dependencies in sequences \cite{6795963}. For a sequence-based experiment, we use CNN as a feature extractor that learns spatial patterns in the range-azimuth feature maps; its output is temporally aggregated by a bidirectional Long Short-Term Memory (LSTM) head, followed by a linear classifier. We stack 10 consecutive frames, which corresponds to 1 second of data, as a chunk with a stride of 1 as visualized as Fig.~\ref{fig:model_archs} (bottom). Table  \ref{tab:rasso_sequence} demonstrates that RASSO improves the accuracy of all datasets, with a narrow high-performance band of $99.4$–$99.6\%$, and both F1 scores $\ge 0.983$. 

\begin{table}[t]
\centering
\caption{RASSO Results – Sequence Based (CNN-LSTM)}
\label{tab:rasso_sequence}
\resizebox{\linewidth}{!}{%
\begin{tabular}{@{} l|l c c c @{}}
\toprule
\textbf{Subject} & \textbf{Variant} & \textbf{Acc. (\%)} & \textbf{F1 (empty)} & \textbf{F1 (1-person)} \\
\midrule

\multirow{2}{*}{\makecell[l]{Subject 1\\(Held-out Test Set)}} 
 & Baseline RA  & 0.9561 & 0.9607 & 0.9502 \\
 & \textbf{RASSO RA} & \textbf{0.9845} & \textbf{0.9857} & \textbf{0.9831} \\
\midrule

\multirow{2}{*}{\makecell[l]{Subject 2\\(Generalization Test Set)}} 
 & Baseline RA  & 0.9772 & 0.9449 & 0.9856 \\
 & \textbf{RASSO RA} & \textbf{0.9948} & \textbf{0.9868} & \textbf{0.9967} \\
\midrule

\multirow{2}{*}{\makecell[l]{Subject 3\\(Generalization Test Set)}} 
 & Baseline RA  & 0.9717 & 0.9522 & 0.9799 \\
 & \textbf{RASSO RA} & \textbf{0.9942} & \textbf{0.9899} & \textbf{0.9960} \\
\midrule

\multirow{2}{*}{\makecell[l]{Subject 4\\(Generalization Test Set)}} 
 & Baseline RA  & 0.9652 & 0.9415 & 0.9752 \\
 & \textbf{RASSO RA} & \textbf{0.9956} & \textbf{0.9921} & \textbf{0.9969} \\

\bottomrule
\end{tabular}%
}
\end{table}

\subsection{Uncertainty Quantification (CNN-LSTM, Subject 2-4)}
\begin{figure}[htb] 
\centering 
\includegraphics[width=\linewidth]{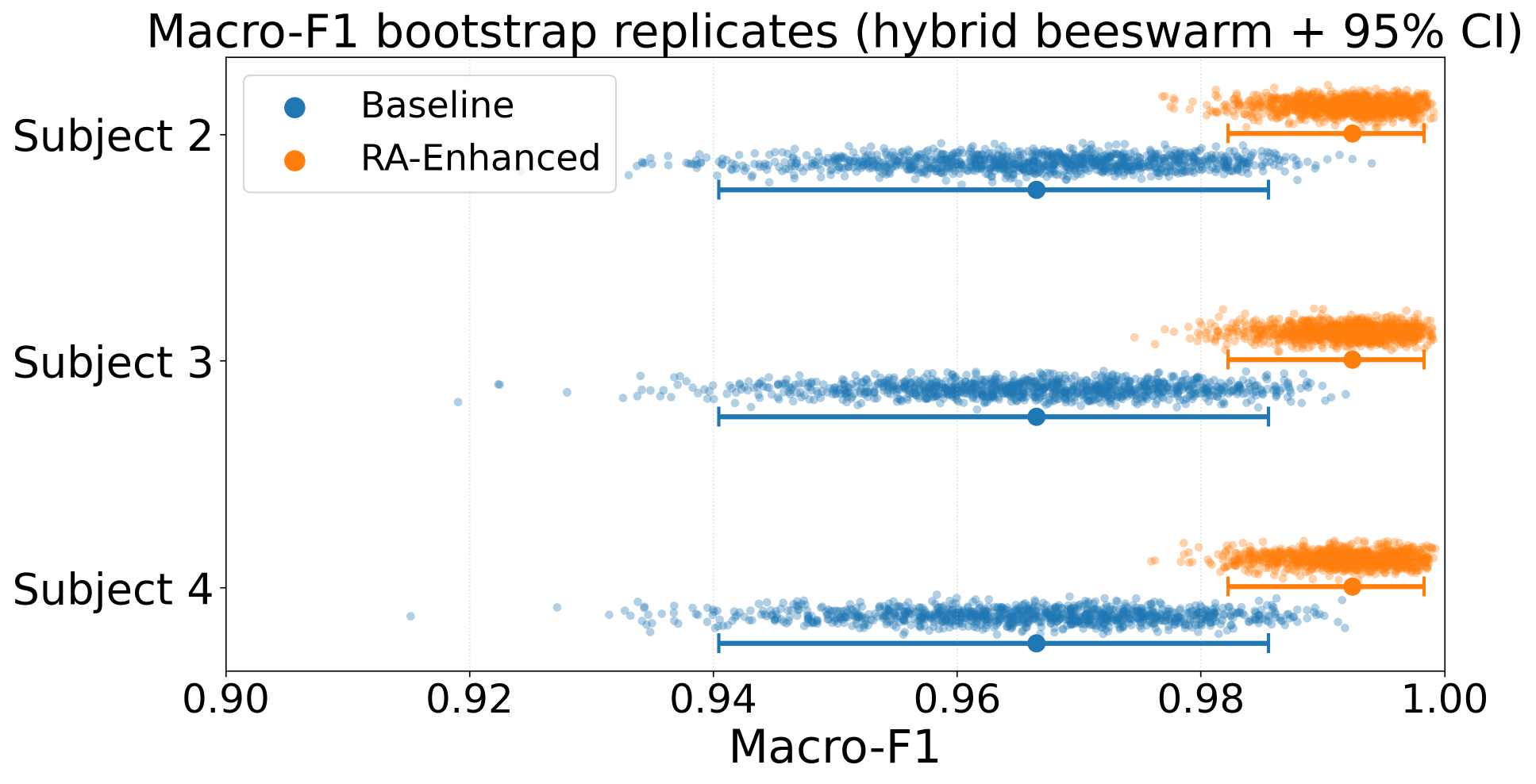}
\caption{Macro-F1 bootstrap replicates (hybrid beeswarm) with 95\% CIs for Subjects 2–4. Each dot is one session-level bootstrap replicate. The solid dot and whiskers are the full-set estimate and its 95\% CI. Orange (RASSO-Enhanced) consistently exceeds blue (Baseline).}
\label{fig:beeswarm_macroF1_withCI}
\end{figure}

For each generalization subject (2–4) and each model (Baseline and RASSO-Enhanced, both trained on Subject~1; see Section~\ref{sub:Data-Driven-Classification}), we draw $B = 1000$ bootstrap replicates. For each replicate $b$, we compute Accuracy, the per-class F1 scores (empty\_room, person\_present), and the macro-averaged F1. We report the point estimate from the full (non-resampled) test set, and a 95\% confidence interval given by the $[2.5, 97.5]$ empirical percentiles of the $B$ bootstrap values. The Fig.~\ref{fig:beeswarm_macroF1_withCI} analyzes the uncertainty of Macro-F1 for these subjects; each point indicates a session-level bootstrap replicate, and the circles and whiskers show the full-set estimate and 95\% confidence interval. RASSO-Enhanced (orange) swarms consistently outperform Conventional (blue) ones across Subjects 2-4, with tighter CIs near unity. As a result, RASSO shifts Macro-F1 upward by $2.6-3.6$ percentage points of the median on generalization subjects, and its 95\% interval is higher than 98\% Macro-F1.

\begin{figure}[htb] 
\centering 
\includegraphics[width=\linewidth]{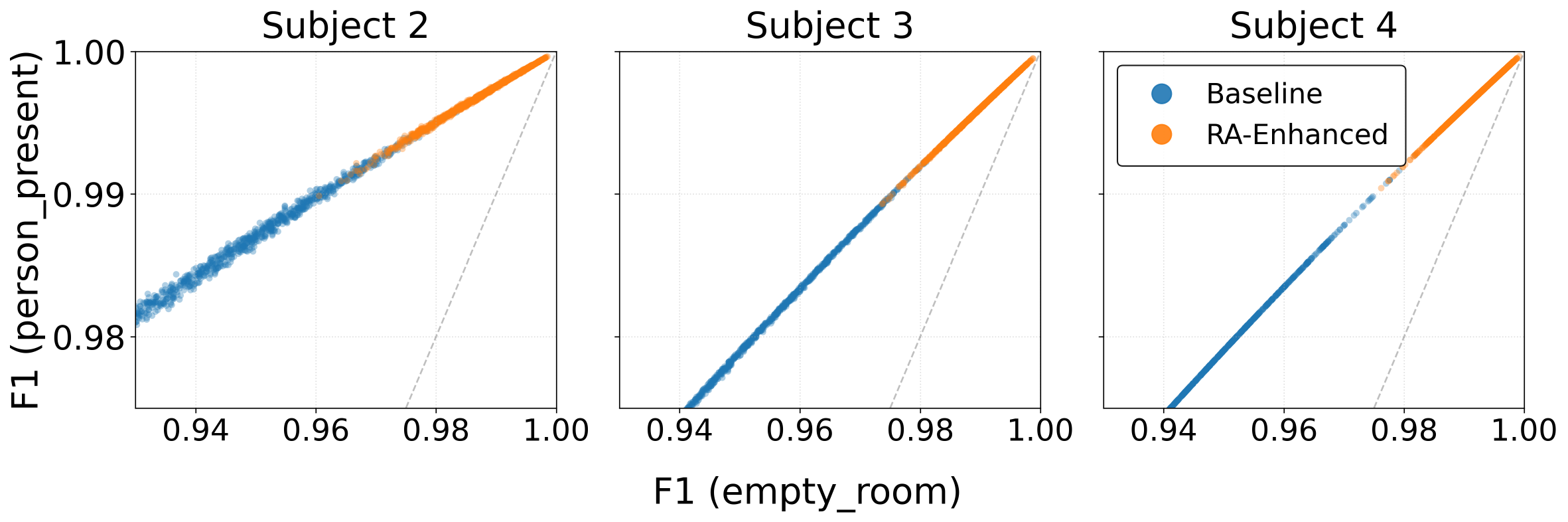}
\caption{Per-class uncertainty. Bootstrap join scatter of $F1_{\text{empty}}(\text{x})$ vs. $F1_{\text{person}}(\text{y})$. The grey diagonal marks equal per-class performance.}
\label{fig:scatter_F1_empty_vs_person_pretty_v2}
\end{figure}
\begin{figure}[htb] 
\centering 
\includegraphics[width=\linewidth]{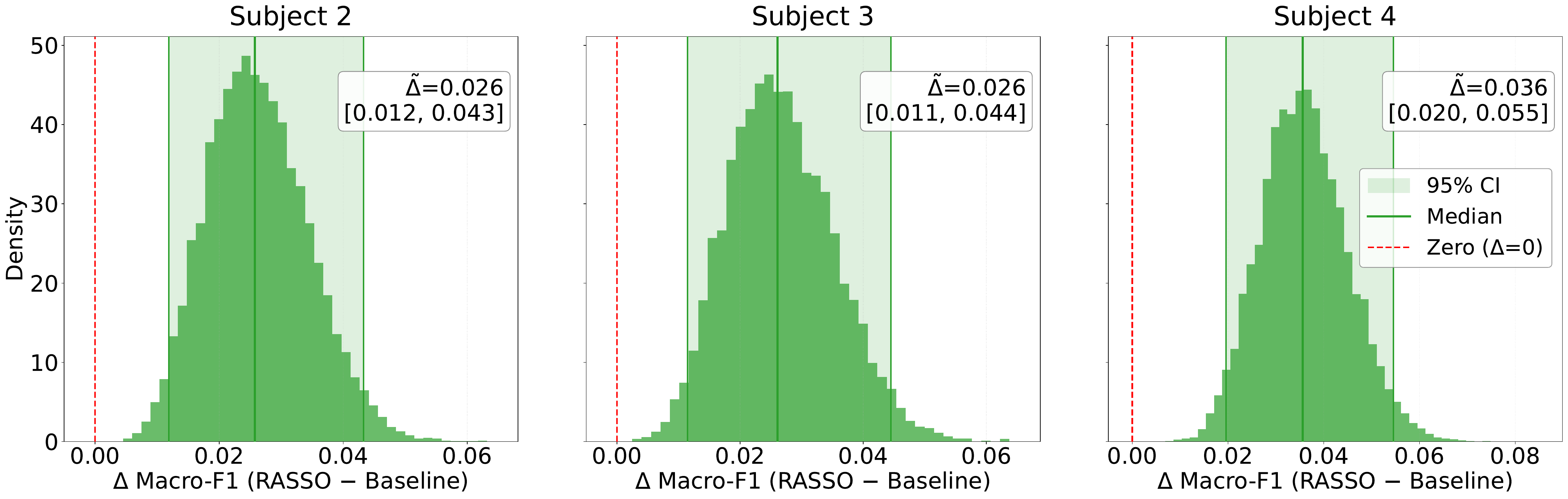}
\caption{Paired improvement distributions. Histograms of $\Delta$ = \text{Macro-$\text{F1}_{\text{RASSO}}$} - \text{Macro-$\text{F1}_{\text{Baseline}}$} with median (solid green), 95\% CI (shaded band), and a red dashed line at zero.}  
\label{fig:delta_macro_hist}
\end{figure}
Fig.~\ref{fig:scatter_F1_empty_vs_person_pretty_v2} presents the per-class behaviour under bootstrap joint scatter of F1 (empty room) versus F1 (person present) for Subjects 2-4. RASSO-Enhanced (orange) concentrates on the upper-right frontier, indicating simultaneous gains in both classes. This suggests that RASSO helps reduce both missed detections in person-present cases and false alarms in empty-room cases. Finally, we compute the paired bootstrap replicate differences for the same sessions for each replicate $r=1...1000$ in the same set, as visualized in Fig.~\ref{fig:beeswarm_macroF1_withCI}.
\begin{equation}
    \Delta_r =  F_{1,macro}^{\text{RASSO}} - F_{1,macro}^{\text{Conv}} .
\end{equation}
Fig.~\ref{fig:delta_macro_hist} presents the distribution of paired differences
$\Delta = \mathrm{Macro}\text{-}\mathrm{F1}_{\mathrm{RASSO}} - \mathrm{Macro}\text{-}\mathrm{F1}_{\mathrm{Conv}}$
computed for each subject. For all three subjects, the 95\% confidence interval lies strictly above zero,
confirming statistically significant improvements (empirical $p \leq 0.001$ with 1000 resamples).
The median and low/high values of the Macro-F1 differences are
0.026 [0.012, 0.043], 0.026 [0.011, 0.044], and 0.036 [0.020, 0.005] for subjects 2, 3, and 4, respectively.

\section{Conclusion}
\label{sec:conclusion}
This paper presented RASSO, 
an invertible Doppler-domain non-linear warping and resampling designed to enhance semi-static human occupancy detection using low-resolution SIMO FMCW radar in realistic long-term care environments. By densifying the Doppler sampling around near-zero velocities prior to Capon beamforming and CA–CFAR detection, RASSO produces range–azimuth maps with higher signal concentration and lower background variance, yielding an average SNR improvement from 6.88 dB to 9.55 dB in representative recordings. Across CFAR-based (knowledge-driven) evaluations, the RASSO-enhanced pipeline achieves the highest detection-curve AUC (0.981) and the best recall at FAR = 1\% and 5\% (0.920 and 0.947) among all compared front-ends, and improves macro-F1 by approximately 6–10 percentage points across subjects in cross-subject CA–CFAR experiments, while maintaining very high F1 for the person-present class. When used as input to lightweight data-driven models, RASSO-based RA maps further enable a SimpleCNN and a CNN–LSTM to reach 95–99\% and 99.4–99.6\% accuracy, respectively, with session-level bootstrap analysis confirming statistically significant macro-F1 gains of 2.6–3.6 percentage points over the conventional non-warped baseline. Future work will extend this framework to more complex and clinically realistic long-term care scenarios in which multiple residents, visitors, and nursing staff may be present and interacting. We also plan to study domain adaptation across facilities and room layouts, and to integrate RASSO with radar-based vital-sign extraction into a unified pipeline that jointly estimates occupancy and physiological indicators, enabling richer health monitoring while preserving the inherent privacy of radar sensing.

\section*{Acknowledgments}
The authors would like to give thanks to ElephasCare, Schlegel-UW Research Institute for Aging, and MITACS for supporting this study. In addition, resources used in preparing this research were provided, in part, by the Province of Ontario, the Government of Canada through CIFAR, and companies sponsoring the Vector Institute \href{www.vectorinstitute.ai/partnerships/}{\url{www.vectorinstitute.ai/partnerships/}}.

%
\bibliographystyle{IEEEtran}
\bibliography{refs}

@INPROCEEDINGS{9951449,
  author={Luo, Yihong and Li, Xiaoxia},
  booktitle={2022 2nd International Conference on Frontiers of Electronics, Information and Computation Technologies (ICFEICT)}, 
  title={Indoor Human Location Method for FMCW Radar Using Standard Deviation Weighting}, 
  year={2022},
  volume={},
  number={},
  pages={159-163},
  doi={10.1109/ICFEICT57213.2022.00036}}

@INPROCEEDINGS{10289261,
  author={Kaiser, Kevin and Stadelmayer, Thomas and Felgentreff, Jens and Weigel, Robert and Lurz, Fabian and Santra, Avik},
  booktitle={2023 20th European Radar Conference (EuRAD)}, 
  title={Improved Indoor Semi-Static Human Target Detection and Localization using FMCW Radar}, 
  year={2023},
  volume={},
  number={},
  pages={298-301},
  doi={10.23919/EuRAD58043.2023.10289261}}

@INPROCEEDINGS{10118759,
  author={Li, Hongchun and Xie, Lili and Zhao, Qian and Tian, Jun and Konno, Takeshi},
  booktitle={2023 IEEE Wireless Communications and Networking Conference (WCNC)}, 
  title={Static Human Localization Using FMCW MIMO Radar}, 
  year={2023},
  volume={},
  number={},
  pages={1-6},
  doi={10.1109/WCNC55385.2023.10118759}}

@ARTICLE{10789192,
  author={Kahya, Sabri Mustafa and Yavuz, Muhammet Sami and Steinbach, Eckehard},
  journal={IEEE Transactions on Radar Systems}, 
  title={HOOD: Real-Time Human Presence and Out-of-Distribution Detection Using FMCW Radar}, 
  year={2025},
  volume={3},
  number={},
  pages={44-56},
  doi={10.1109/TRS.2024.3514840}}

@ARTICLE{8703820,
  author={Will, Christoph and Vaishnav, Prachi and Chakraborty, Abhiram and Santra, Avik},
  journal={IEEE Sensors Journal}, 
  title={Human Target Detection, Tracking, and Classification Using 24-GHz FMCW Radar}, 
  year={2019},
  volume={19},
  number={17},
  pages={7283-7299},
  doi={10.1109/JSEN.2019.2914365}}

@Article{bios15050273,
AUTHOR = {Song, Chenyan and Yavari, Ehsan and Gao, Xiaomeng and Lubecke, Victor M. and Boric-Lubecke, Olga},
TITLE = {Respiration Signal Pattern Analysis for Doppler Radar Sensor with Passive Node and Its Application in Occupancy Sensing of a Stationary Subject},
JOURNAL = {Biosensors},
VOLUME = {15},
YEAR = {2025},
NUMBER = {5},
ARTICLE-NUMBER = {273},
URL = {https://www.mdpi.com/2079-6374/15/5/273},
PubMedID = {40422012},
ISSN = {2079-6374},
DOI = {10.3390/bios15050273}
}

@misc{infineon2024bgt60tr13c,
  author       = {{Infineon Technologies AG}},
  title        = {BGT60TR13C 60\,GHz Radar Sensor},
  howpublished = {\emph{Infineon Product Page}},
  month        = {March},
  year         = {2024},
  note         = {[Online]. Available: \url{https://www.infineon.com/cms/en/product/sensor/radar-sensors/radar-sensors-for-iot/60ghz-radar/bgt60tr13c/} [Accessed: April 15, 2025]}
}

@ARTICLE{6795963,
  author={Hochreiter, Sepp and Schmidhuber, Jürgen},
  journal={Neural Computation}, 
  title={Long Short-Term Memory}, 
  year={1997},
  volume={9},
  number={8},
  pages={1735-1780},
  doi={10.1162/neco.1997.9.8.1735}}

@ARTICLE{10012054,
  author={Abedi, Hajar and Ansariyan, Ahmad and Morita, Plinio P. and Wong, Alexander and Boger, Jennifer and Shaker, George},
  journal={IEEE Internet of Things Journal}, 
  title={AI-Powered Noncontact In-Home Gait Monitoring and Activity Recognition System Based on mm-Wave FMCW Radar and Cloud Computing}, 
  year={2023},
  volume={10},
  number={11},
  pages={9465-9481},
  doi={10.1109/JIOT.2023.3235268}}

@INPROCEEDINGS{9704291,
  author={Abedi, Hajar and Ansariyan, Ahmad and Morita, Plinio P and Boger, Jennifer and Wong, Alexander and Shaker, George},
  booktitle={2021 IEEE International Symposium on Antennas and Propagation and USNC-URSI Radio Science Meeting (APS/URSI)}, 
  title={Sequential Deep Learning for In-Home Activity Monitoring Using mm-Wave FMCW Radar}, 
  year={2021},
  volume={},
  number={},
  pages={1499-1500},
  doi={10.1109/APS/URSI47566.2021.9704291}}

@INPROCEEDINGS{10784889,
  author={Abedi, Hajar and Ansariyan, Ahmad and Shaker, George},
  booktitle={2024 IEEE SENSORS}, 
  title={Contactless In-Bed Detection Using a Low-Cost Low-Resolution Radar}, 
  year={2024},
  volume={},
  number={},
  pages={1-4},
  doi={10.1109/SENSORS60989.2024.10784889}}

@INPROCEEDINGS{10880536,
  author={Visser, Joshua and Abedi, Hajar and Karmani, Sachin and Shaker, George},
  booktitle={2025 IEEE MTT-S Latin America Microwave Conference (LAMC)}, 
  title={AI based Activity Monitoring in Washrooms using Low-Resolution Radar}, 
  year={2025},
  volume={},
  number={},
  pages={78-81},
  doi={10.1109/LAMC63321.2025.10880536}}

@book{book,
author = {Santra, Avik and Hazra, Souvik},
year = {2020},
month = {09},
pages = {},
title = {Deep Learning Applications of Short-Range Radars},
isbn = {9781630817466}
}

@article{Miller2009FundamentalsOR,
  title={Fundamentals of Radar Signal Processing (Richards, M.A.; 2005) [Book review]},
  author={Robert Miller},
  journal={IEEE Signal Processing Magazine},
  year={2009},
  volume={26},
  pages={100-101},
  url={https://api.semanticscholar.org/CorpusID:35286446}
}

@INPROCEEDINGS{10289281,
  author={Nguyen, Minh Q. and Feger, Reinhard and Wagner, Thomas and Stelzer, Andreas},
  booktitle={2023 20th European Radar Conference (EuRAD)}, 
  title={Analysis of 2D CA-CFAR for DDMA FMCW MIMO Radar}, 
  year={2023},
  volume={},
  number={},
  pages={423-426},
  doi={10.23919/EuRAD58043.2023.10289281}}

@article{Abedi2020OnTU,
  title={On the Use of Low-Cost Radars and Machine Learning for In-Vehicle Passenger Monitoring},
  author={Hajar Abedi and Shenghang Luo and George Shaker},
  journal={2020 IEEE 20th Topical Meeting on Silicon Monolithic Integrated Circuits in RF Systems (SiRF)},
  year={2020},
  pages={63-65},
  url={https://api.semanticscholar.org/CorpusID:214595336}
}

@techreport{InfineonAN155322,
  author       = {{Infineon Technologies AG}},
  title        = {Digital Beamforming Using the Demo {BGT60TR13C} Radar Sensor},
  type         = {Application Note},
  number       = {AN155322},
  institution  = {Infineon Technologies AG},
  year         = {2023},
  note         = {Rev. per vendor site},
  url          = {https://www.infineon.com/assets/row/public/documents/24/42/infineon-an155322-digital-beamforming-using-the-demo-bgt60tr13c-radar-sensor-applicationnotes-en.pdf},
  urldate      = {2025-10-16}
}

@techreport{infineon_an141319,
  author       = {Infineon Technologies AG},
  title        = {Ceiling-mounted occupancy detection using XENSIV\textsuperscript{\texttrademark} DEMO BGT60TR13C 60~GHz radar},
  institution  = {Infineon},
  type         = {Application Note AN141319},
  number       = {AN141319},
  year         = {2024},
  month        = {Feb},
  note         = {Available: \url{https://www.infineon.com/assets/row/public/documents/24/42/infineon-an141319-ceiling-mounted-occupancy-detection-using-xensiv-demo-bgt60tr13c-60-ghz-radar-applicationnotes-en.pdf}}
}

@INPROCEEDINGS{11031826,
  author={Fard, Ali Samimi and Mashhadigholamali, Mohammadreza and Zolfaghari, Samaneh and Abedi, Hajar and Chakraborty, Mainak and Karmani, Sachin and Borzì, Luigi and Daneshtalab, Masoud and Shaker, George},
  booktitle={2025 IEEE International Radar Conference (RADAR)}, 
  title={Fall Detection in Ambient-Assisted Living Environments Using FMCW Radars and Deep Learning}, 
  year={2025},
  volume={},
  number={},
  pages={1-6},
  doi={10.1109/RADAR52380.2025.11031826}}

@ARTICLE{995829,
  author={Umesh, S. and Cohen, L. and Nelson, D.},
  journal={IEEE Signal Processing Letters}, 
  title={Frequency warping and the Mel scale}, 
  year={2002},
  volume={9},
  number={3},
  pages={104-107},
  doi={10.1109/97.995829}}

@misc{TI_FMCW_Training,
  title        = {mmWave Sensing: FMCW Radar Overview (training slides)},
  author       = {{Texas Instruments}},
  howpublished = {\url{https://www.ti.com/video/series/mmwave-training-series.html}},
  year         = {2024},
}

@misc{Infineon_DSP_Handout,
  title        = {FMCW Radar --- Digital Signal Processing},
  author       = {{Infineon Technologies}},
  howpublished = {\url{https://www.infineon.com/assets/row/public/documents/24/56/infineon-fmcw-radar-digital-signal-processing-handout-training-en.pdf}},
  year         = {2022},
  note         = {De-chirp, range/Doppler processing chain, RDC organization}
}

@phdthesis{phdthesis,
author = {Gao, Xiangyu},
year = {2021},
month = {12},
pages = {},
title = {Towards Millimeter-wave Radar Signal Processing and Learning-Based Application}
}

@ARTICLE{10554983,
  author={Kong, Hao and Huang, Cheng and Yu, Jiadi and Shen, Xuemin},
  journal={IEEE Communications Surveys \& Tutorials}, 
  title={A Survey of mmWave Radar-Based Sensing in Autonomous Vehicles, Smart Homes and Industry}, 
  year={2025},
  volume={27},
  number={1},
  pages={463-508},
  doi={10.1109/COMST.2024.3409556}}

@article{DUHAMEL1990259,
title = {Fast fourier transforms: A tutorial review and a state of the art},
journal = {Signal Processing},
volume = {19},
number = {4},
pages = {259-299},
year = {1990},
issn = {0165-1684},
doi = {https://doi.org/10.1016/0165-1684(90)90158-U},
url = {https://www.sciencedirect.com/science/article/pii/016516849090158U},
author = {P. Duhamel and M. Vetterli}
}

@misc{kahya2023mcroodmulticlassradaroutofdistribution,
      title={MCROOD: Multi-Class Radar Out-Of-Distribution Detection}, 
      author={Sabri Mustafa Kahya and Muhammet Sami Yavuz and Eckehard Steinbach},
      year={2023},
      eprint={2303.06232},
      archivePrefix={arXiv},
      primaryClass={cs.CV},
      url={https://arxiv.org/abs/2303.06232}, 
}
\newpage
\section*{Biography Section}
\begin{IEEEbiography}[{\includegraphics[width=1in,height=1.25in,clip,keepaspectratio]{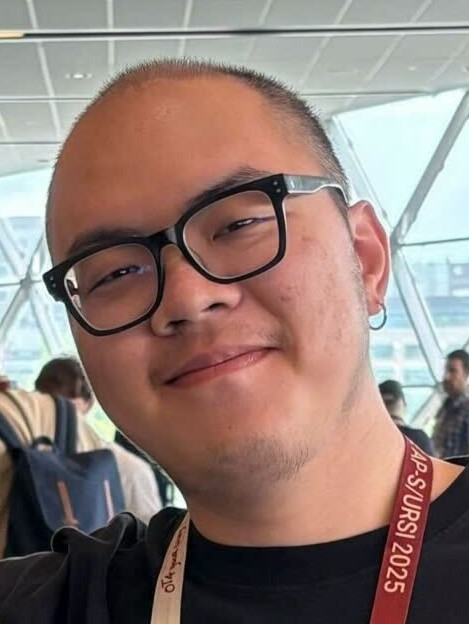}}]{Huy Trinh}
received the B.Sc.\ degree in science and engineering (with a major in embedded systems and mathematics) from Tampere University, Tampere, Finland, in 2023. He is currently a MASc candidate in electrical and computer engineering at the University of Waterloo, Waterloo, ON, Canada, where he is affiliated with the Wireless Sensors and Devices Laboratory (WSDL) and the Socially Embedded Machine Intelligence (SEMI) Laboratory. His research spans mmWave FMCW radar sensing, signal processing, and machine learning for indoor human presence detection and activity recognition. He was previously an intern at CSC (Finland), CINECA (Italy) and RIKEN (Japan). 
\end{IEEEbiography}
\begin{IEEEbiography}[{\includegraphics[width=1in,height=1.25in,clip,keepaspectratio]{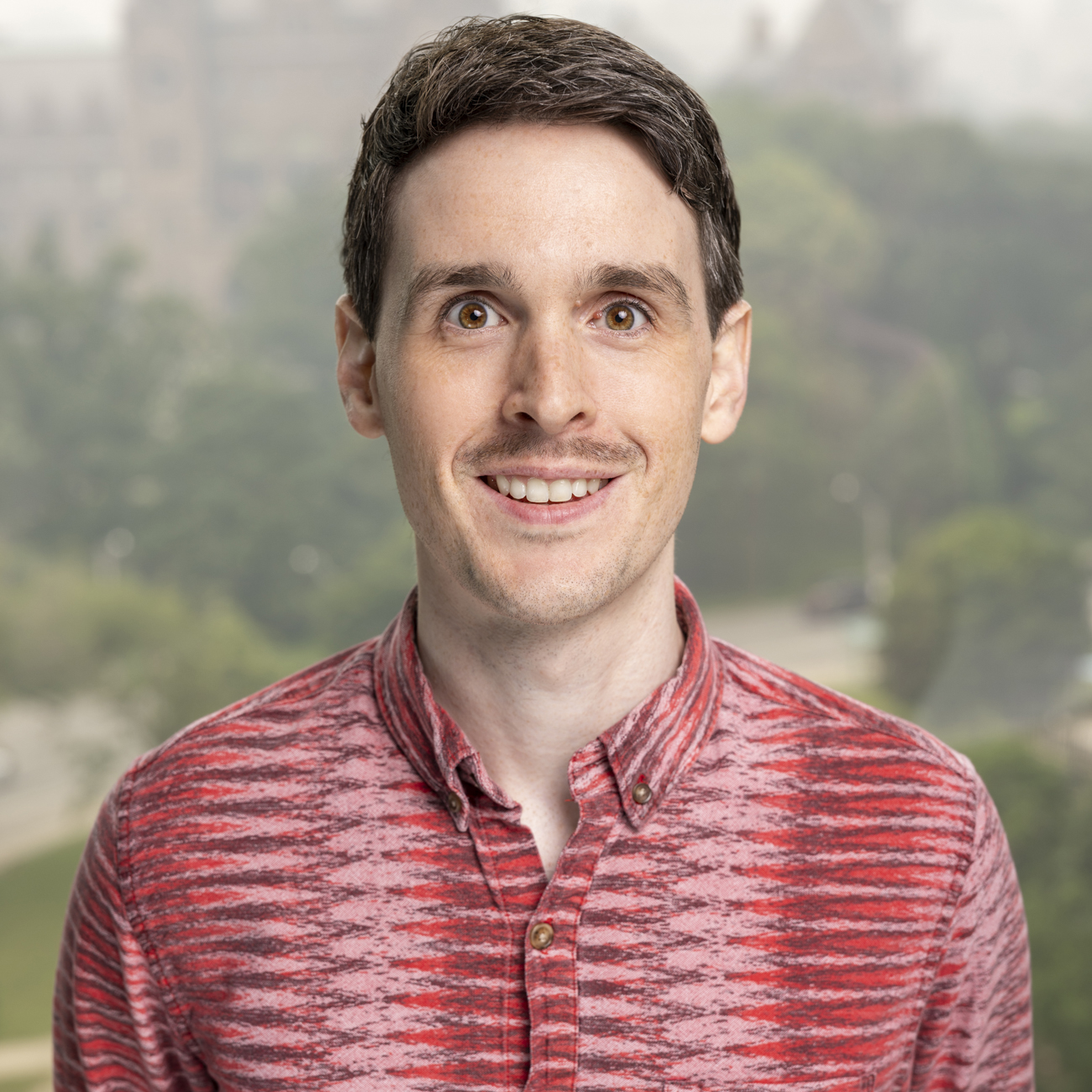}}]{Elliot Creager}
is an Assistant Professor of Electrical and Computer Engineering at the University of Waterloo, where he is the director of the Socially Embedded Machine Intelligence (SEMI) Lab. His research focuses on building reliable and equitable AI systems that promote user data autonomy. Prof.\ Creager is a Faculty Affiliate of the Vector Institute for Artificial Intelligence and the Schwartz Reisman Institute for Technology and Society. He is also cross-appointed to the Department of Management Science and Engineering and the David R.\ Cheriton School of Computer Science, and is a faculty member of the Waterloo Data and Artificial Intelligence Institute and the Waterloo Cybersecurity and Privacy Institute. He completed his Ph.D.\ at the University of Toronto in 2023 and was previously an intern and student researcher at Google Brain in Toronto.
\end{IEEEbiography}
\begin{IEEEbiography}[{\includegraphics[width=1in,height=1.25in,clip,keepaspectratio]{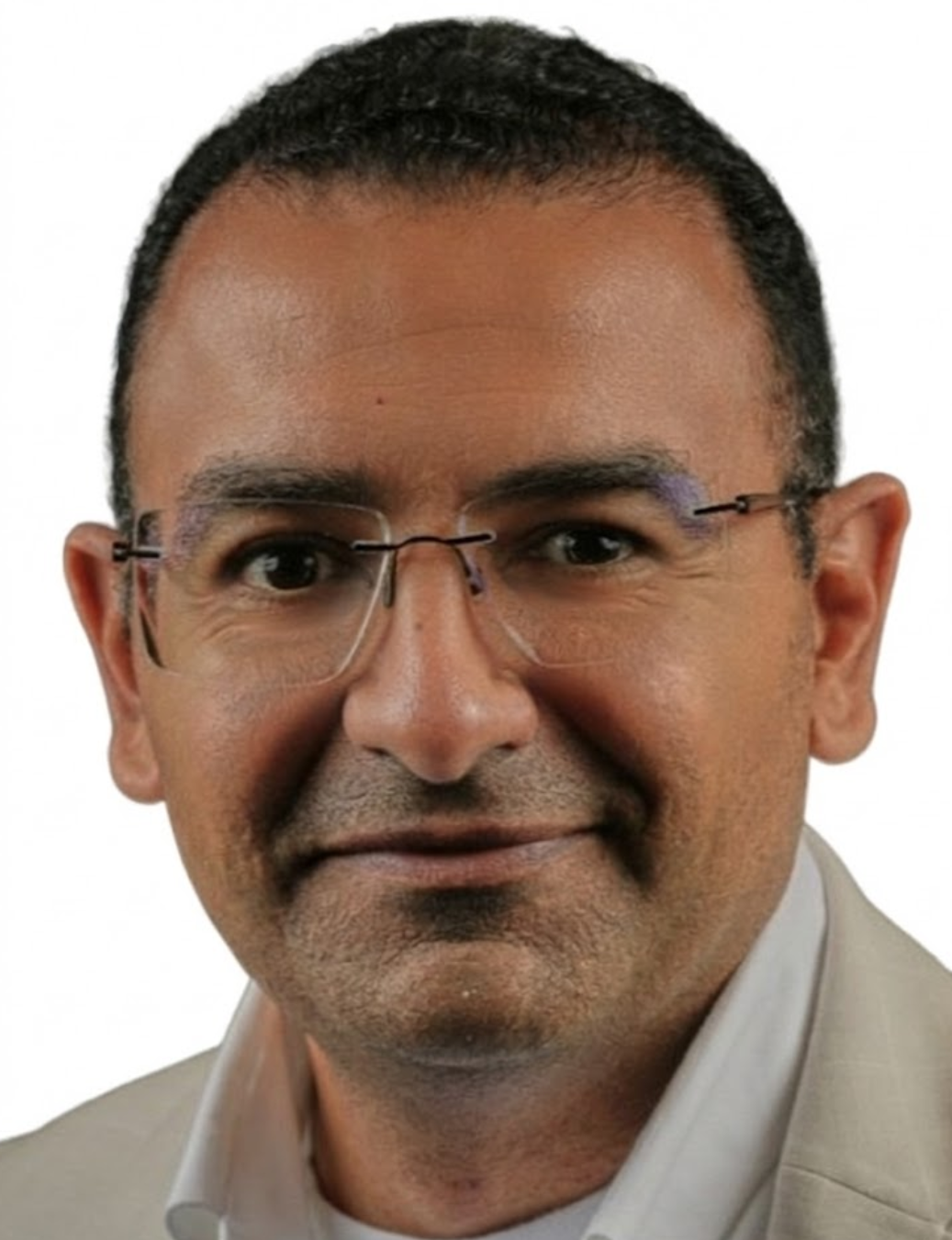}}]{George Shaker}
(S'97--M'11--SM'15) is the lab director of the Wireless Sensors and Devices Laboratory at the University of Waterloo, where he is an Adjunct Associate Professor in the Department of Electrical and Computer Engineering. Previously, he was an NSERC scholar at the Georgia Institute of Technology. Dr.\ Shaker also held multiple roles with Research In Motion (BlackBerry). He is the Chief Scientist at Spark Tech Labs, which he co-founded in 2011. With over twenty years of industrial experience in technology and more than ten years as a faculty member leading projects related to the application of wireless sensor systems for healthcare, automotive, and unmanned aerial vehicles, Prof.\ Shaker has made many design contributions in commercial products available from startups and multinationals. A sample list includes Google, COM DEV, Honeywell, BlackBerry, Spark Tech Labs, Bionym, Lyngsoe Systems, ON Semiconductors, Ecobee, Medella Health, NERV Technologies, Novela, Thalmic Labs, North, General Dynamics Land Systems, General Motors, Toyota, Maple Lodge Farms, Rogers Communications, and Purolator. He is currently an IEEE AP-S Distinguished Industry Speaker and an IEEE Sensors Council Distinguished Lecturer.

Dr.\ Shaker has authored or coauthored more than 200 publications and over 35 patents and patent applications. He has received multiple recognitions and awards, including the IEEE AP-S Best Paper Award, several IEEE AP-S Honorable Mention Best Paper Awards, the IEEE Antennas and Propagation Graduate Research Award, the IEEE MTT-S Graduate Fellowship, the Electronic Components and Technology Best of Session paper award, and the IEEE Sensors most popular paper award. Several papers he coauthored in IEEE journals were among the top 25 downloaded papers on IEEE Xplore for several consecutive months. He was the supervisor of the student team winning the third best design contest at IEEE AP-S 2016 and 2025, and coauthor of the ACM MobileHCI 2017 Best Workshop Paper Award and the 2018 Computer Vision Conference Imaging Best Paper Award. He co-received with his students several research recognitions, including the NSERC Top Science Research Award 2019, IEEE AP-S HM Paper Awards (2019, 2022, 2023, 2024, and 2025), Biotec Top Demo Award 2019, arXiv Top Downloaded Paper (medical device category) 2019, Velocity Fund 2020, NASA Tech Briefs HM Award (medical device category) 2020, UW Concept 2021, UK Dragons Canadian Competition 2021, CMC Nano 2021, COIL COLAB 2022, Wiley Engineering Reports Top Downloaded Paper for 2022, Canadian Space Agency Cubesat Design winner 2023, IEEE NEMO Best Paper Award 2024, Nature Communications Engineering Top 25 Downloaded Papers in 2024, IEEE MAPCON Best Paper Award 2024, and iWAT 2025 Paper Finalist.
\end{IEEEbiography}

\vfill
\end{document}